\documentclass[twocolumn]{aastex63}
\usepackage[cmex10]{amsmath}
\usepackage{amssymb}    

\newcommand{\asec}{$^{\prime\prime}$}
\newcommand{\amin}{$^{\prime}$}
\newcommand*{\chg}{}
\newcommand*{\chga}{}
\received{January 17, 2025}
\revised{March 24, 2025}
\revised{April 11, 2025}
\accepted{April 12, 2025}
\shorttitle{MeerKAT MW Bulge}
\shortauthors{Cotton et al.}

\begin{document}

\title{MeerKAT 1.3 GHz Observations Towards the Milky Way Bulge}

\correspondingauthor{William Cotton}
\email{bcotton@nrao.edu}
\author[0000-0001-7363-6489]{W.~D.~Cotton}
\affiliation{National Radio Astronomy Observatory \\
520 Edgemont Road \\
Charlottesville, VA 22903, USA}
\affiliation{South African Radio Astronomy Observatory \\
Liesbeek House, River Park, Gloucester Road \\
Cape Town, 7700, South Africa} 

\author[0009-0000-0057-992X]{P.~J.~Agnihotri }
\affiliation{National Radio Astronomy Observatory \\
520 Edgemont Road \\
Charlottesville, VA 22903, USA}
\affiliation{San Francisco State University \\
San Francisco CA, USA} 

\author[0000-0002-1873-3718]{F.~Camilo}
\affiliation{South African Radio Astronomy Observatory \\
Liesbeek House, River Park, Gloucester Road \\
Cape Town, 7700, South Africa}

\author[0000-0003-3272-9237]{E. Polisensky}
\affiliation{U.S.\ Naval Research Laboratory,  4555 Overlook Ave SW,  Washington,  DC 20375,  USA}

\author[0009-0006-5070-6329]{S. D. Hyman}
\affiliation{Department of Engineering and Physics, Sweet Briar College, Sweet Briar, VA 24595, USA}



\begin{abstract}
We present a MeerKAT survey of portions of the Milky Way bulge.
The survey covers 172.8 square degrees in two contiguous mosaics above and below the Galactic Center as well as 32 single pointing fields at higher longitudes.
The resolution of the images is $\sim$8\asec\ at a frequency of 1333 MHz with a typical Stokes I RMS of 20 $\mu$Jy Beam$^{-1}$.
Most of the emission seen is from background extragalactic sources but
many compact Galactic objects are identifiable by their polarization
properties. 
Apparent polarized emission resulting from fine scale Faraday rotation
in the ISM is widespread in this region of the Galaxy.
The survey is used to search for background Giant Radio Galaxies,  $>$700 kpc in size, identifying 17 such objects.
Data products include FITS images of Stokes I, Q, U and V as well as a Faraday analysis and lists of compact total intensity and polarized sources.
\end{abstract}



\section{Introduction} 
The Galactic center of our galaxy currently has a fairly low level of
activity but in the last few million years an explosive event has
occurred, either an outburst of the supermassive black hole (SMBH),
Sgr A*, or a burst of massive star formation
\citep{FermiBubble,MKBubble}.  
One  of the remnants of this episode is the pair of ``MeerKAT
Bubbles'' \citep{MKBubble} that appear to be emerging perpendicular to
the Galactic plane from the Galactic center.
Further out are the ``Fermi Bubbles'' \citep{FermiBubble} thought to
  be the result of an even earlier event.
Since the Galaxy is largely optically thin at radio wavelengths, it is
an ideal region of the spectrum to search for the remnants of other
explosive events in the Galactic center.
The region above and below the Galactic center, i.e. in the halo and
bulge have not previously been surveyed with adequate resolution and
sensitivity in the radio to have detected other, similar features.

In this paper we describe observations of the region above and below the
Galactic center, the bulge and halo, which could detect the remnants of
previous outbursts at the Galactic center.
The region in the immediate vicinity of the Galactic center, including
the MeerKAT Bubbles is covered in \cite{Heywood2022} and
\cite{MKBubble}.
A MeerKAT survey of the Galactic plane is reported in \cite{MKGPLS}. 

The Milky Way bulge is composed of an old population of stars, 10
billion years old \citep{bulge} and is devoid of recent star formation.
However, the line of sight through the bulge goes through a
significant distance in the Galactic plane so observations in that
direction are expected to include objects from the disk population as
well as pulsars and other objects in the bulge.
The dataset described here has been used to search for pulsars
\citep{Frail2024}.  

The observations are described in Section \ref{Obs}, the data
processing in Section \ref{DataProc}, limitations of the data are
described in Section \ref{Limitation} and the data products available
in Section \ref{Products}.
The results of the survey are described in Section \ref{Results} and a
search for Giant Radio Galaxies (GRGs) in the background extragalactic
sky is outlined in Section \ref{GRGs}.
A final summary is given in Section \ref{Summary}.

\section{Observations\label{Obs}}
Observations were made of selected regions of the bulge of the
Milky Way at L band (856 to 1712 MHz) using the MeerKAT array in South
Africa (\cite{Jonas2016}, \cite{Camilo2018}, \cite{DEEP2}.
These used at least 60 of the 64 antenna, 4096 spectral channels
across the band, 8 second integrations and all four of the products of
the orthogonal linearly polarized feeds.
The project code used was SSV-20180505-FC-01 and 419 hours were observed for this project.

\begin{figure}
\centerline{\includegraphics[width=4.0in,angle=-90]{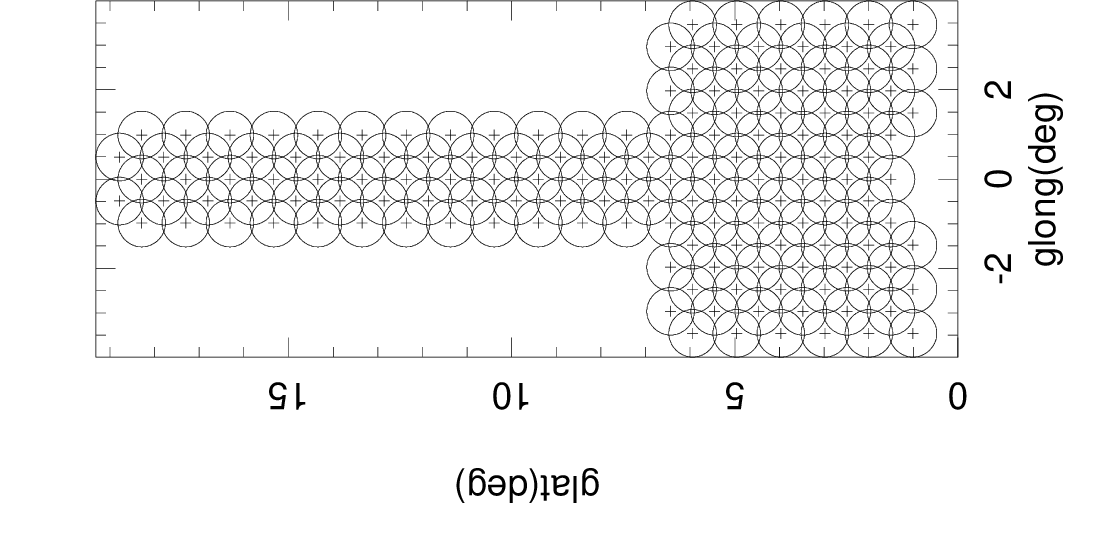}}
\caption{ 
Locations of the pointing centers at positive Galactic latitudes are
shown by pluses and circles show the half power beam sizes at 1.3 GHz.
There is a symmetric coverage at negative latitudes.
}
\label{MosaicCoverageFig}
\end{figure}

There were two components of this project:
\begin{itemize}
\item {\bf Mosaic.}
 Mosaics were made of variable width columns above and below the
Galactic plane, centered at the Galactic center.
These are 8$^\circ$ wide in longitude at the base and 3$^\circ$ wide
at the top and extend from 1$^\circ$ to 19$^\circ$  and -1$^\circ$ to
-19$^\circ$ in latitude.
These regions were observed in 42 sessions (several divided onto
multiple days) with 8 or 9 pointings observed in each of the 8--10
hour sessions.
The observations were made between 2019-12-26 and 2020-08-05.
Observations cycled among several minute scans of each pointing and
calibrators. 
This gives approximately an hour total exposure per pointing but
spread over many hours to improve the uv coverage.
The observing pattern north of the Galactic plane is given in Figure
\ref{MosaicCoverageFig}; a symmetric region was observed south of the plane.
\item {\bf Single Pointings Sample.\label{wider}}
A series of single pointing images was made of a sample of
latitude and longitude locations given in Table \ref{tab:offsets}.
These were observed in 8 $\times$ 5 hour sessions cycling among 4 pointings
and calibrators for approximately 1 hour total exposure per pointing
spread over the 5 hours.
These observations occurred between 2020-12-19 and 2021-01-24.
\end{itemize}

\begin{table}[h]
  \centering
  \begin{tabular}{c|c|c|c|c|c|c|c|c}
   l  $^\circ$ &b  $^\circ$ &b  $^\circ$ &b  $^\circ$ &b  $^\circ$ &b
   $^\circ$ &b  $^\circ$ &b  $^\circ$ &b  $^\circ$ \\ 
    \hline
    15 &  -20& -15& -10&  -5&   5&  10&  15&  20 \\
    20 &  -20& -15& -10&  -5&   5&  10&  15&  19\tablenotemark{a} \\
    40\tablenotemark{b} &  -20& -15& -10&  -5&                 \\
    45 &  -20& -15& -10&  -5&   5&  10&  15&  20 \\
   340\tablenotemark{c} &     &     &    &   &   5&  10&  15&  20 \\
  \hline
\end{tabular}
\caption{Additional Pointings.}
\tablenotetext{a}{b=19 was observed as 3C353 is near the field of b=20.}
\tablenotetext{b}{positive latitudes were not observed.}
\tablenotetext{c}{negative latitudes were not observed.}
\label{tab:offsets}
\end{table}

\section{Data Processing\label{DataProc}}
Processing followed the general approach of \cite{DEEP2},
\cite{XGalaxy}, and \cite{MK_SMC} and used the Obit
package\footnote{http://www.cv.nrao.edu/$\sim$bcotton/Obit.html}
\citep{OBIT}.

\subsection{Calibration}
Calibration and flagging of the data were as described in \cite{DEEP2}
and \cite{MK_SMC}. 
The photometric/bandpass calibrator was PKS~B1934$-$638, the polarization
calibrator was 3C286 and a nearby astrometric calibrator was used in
each session.
The flux density scale was set by the \cite{Reynolds94} spectrum of
PKS~B1934$-$638:
$$
  \log(S) = -30.7667 + 26.4908 \log\bigl(\nu\bigr)
  - 7.0977 \log\bigl(\nu\bigr)^2 $$
  $$+0.605334 \log\bigl(\nu\bigr)^3,$$
where $S$ is the flux density (Jy) and $\nu$ is the frequency (MHz).

\subsection{Imaging}
Imaging of the individual pointing observations in Stokes I, Q, U and
V was as described in \cite{XGalaxy} and \cite{MK_SMC} and used Obit
task MFImage \citep{Cotton2018}. 
Imaging used 14 constant fractional bandwidth (5\%) frequency bins
which are imaged independently but CLEANed jointly to account for the
frequency variations of the sky brightness and antenna gain.
Faceting was used to account for the curvature of the sky.
Images fully covered the sky to a radius of 1.2$^\circ$ with outlying
facets to 1.5$^\circ$ centered on sources estimated to appear in
excess of 1 mJy from the SUMMS \citep{mau03} or NVSS \citep{NVSS} catalogs.
Two iterations of phase only self calibration were always used as was
amplitude and phase self calibration as needed.
CLEANing proceeded to a depth of 100 $\mu$Jy/beam or a maximum of
500,000 components in Stokes I; 40 $\mu$Jy/beam or 50,000
components in Stokes Q and U, and 20 $\mu$Jy/beam or 1,000
components in Stokes V.

\subsection{Peeling\label{peel}}
Some pointings contained sources bright enough that residual
artifacts from the direction independent self calibration
significantly reduced the dynamic range.
Some of these were reimaged using ``Peeling" \citep{Noordam2004} to
reduce these artifacts. 
The emission from sources other than those to be peeled were
subtracted from the visibility data and the peel source was then self
calibrated. 
The response of the peel source was subtracted from the visibility
data using the gain solutions derived for it.
The CLEAN model of the peeled source was restored in the final image.

\subsection{Mosaic Formation}
Individual pointings in the regions covered by the mosaics were
combined in a linear mosaic as described in \cite{SMGPSarXiv} where
primary beam corrections and any astrometric corrections are applied
in the combination.
The pointing images were convolved to an 8\asec\ restoring beam before
combination to assure a constant resolution.
The conversion from Equatorial to Galactic coordinates was made in the
mosaic formation using a simple (not dependent on observing date)
transformation. 
Mosaic cubes were made in Stokes I, Q, U and V.
The primary beam shape used in the correction is that described in
\cite{DEEP2}. 

\subsection{Astrometric Correction}
The astrometric accuracy of the images is limited by a number of
effects, the most significant of which is thought to be the low
accuracy geometric model used in the MeerKAT correlator.
These effects can result in systematic errors up to of order one
arcsecond and are discussed in more detail in \cite{GCLS}.

The Galactic bulge is a region with high optical opacity which reduces
the number of accurate optical/IR positions of background galaxies
which can be used to correct the images.
The MORX catalog \citep{MORX}  of optical/IR positions of radio and X
ray sources was used to determine astrometric corrections for the
mosaics. 
A list of the small, isolated radio sources derived from the initial
version of the mosaic images was compared with entries in the MORX
catalog.
While the centroids in the radio and optical/IR images may not be
exactly concentric, there should not be a systematic offset in the MORX
positions. 

\begin{figure}
\centerline{\includegraphics[width=3.0in,angle=-90]{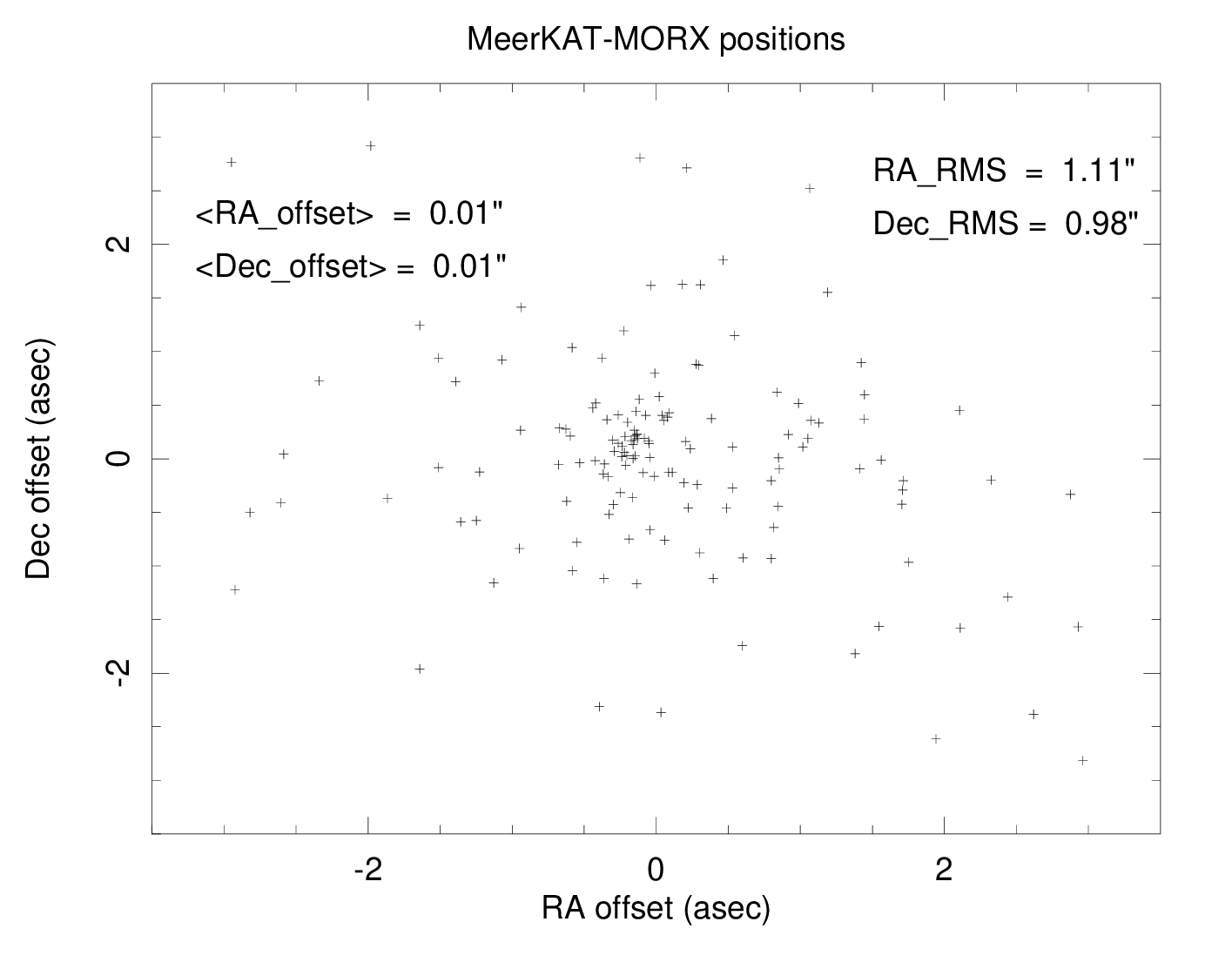}}
\caption{ Residual position offsets between MeerKAT and MORX positions
  after astrometric corrections.
}
\label{fig:astrometry}
\end{figure}

Astrometric errors in the MeerKAT pointing images are highly
correlated, and are corrected in the mosaic formation on an individual
pointing basis. 
Corrections were determined for each pointing included in the combined
mosaics. 
The corrections applied were the weighted averages of the offsets
between the entries in the compact, isolated radio source list and MORX
sources within 5 degrees of the center of the pointing image.  
MORX/MeerKAT matches were required to be within 4\asec\ and the weighting
was inversely proportional to the distance between the pointing center
and each source compared.
These corrections do not exceed 0.6\asec\ in either dimension.
After the corrected mosaics were formed, the comparison between the
MeerKAT and MORX positions were repeated.
This comparison is shown in Figure \ref{fig:astrometry}; the corrected
average offsets were 0.01\asec\ in both Right Ascension and Declination
and the standard deviations of the population were approximately 1\asec.
The astrometric correction was only applied to the mosaics and not the single pointing sample discussed in Section \ref{wider} as there were too few reference MORX positions to make robust corrections.

\subsection{Effective Frequency and Spectral Index}
The effective frequency of the broadband image depends on the details
of how the sub-band images were combined.
The bulk of the pixels in the resultant images have insufficient
signal-to-noise ratio to fit for a spectrum so the combination of the
sub-band images followed that in \cite{SMGPSarXiv,MK_SMC}.
The combination of sub-bands used a weighted average where the
weighting of each sub-band was proportional to the average over all
mosaic images of 1/$\sigma$ where $\sigma$ is the off--source standard
deviation of the pixel values.
1/$\sigma$ is used rather than the more conventional 1/$\sigma^2$
because the steep spectrum of the Galactic background causes the
latter weighting to essentially remove the effects of the lower
frequency sub-bands.
The effective frequency of the broadband mosaics is 1333.1295 MHz.
The same weighting and effective frequency were also applied to the
single pointing images described in Section \ref{wider}.
Spectral indices were fitted in pixels with adequate signal and a
broadband Stokes I brightness in excess of 200 $\mu$Jy/beam.

\subsection{Faraday Analysis}
The Q and U cubes were subjected to a search in Faraday depth in each
pixel to determine the peak polarized intensity, rotation measure and
intrinsic polarization angle.
Much of the polarized emission does not correspond to detected Stokes
I features but is produced by fine scale structure in the magnetized
plasma in the ISM Faraday rotating the smooth polarized emission from the
Galactic disk. 
This fine scale structure is not filtered out by the interferometer array the way the total intensity emission is.

The search range in RM was $\pm$150 rad m$^{-2}$ with an increment of
0.5 rad m$^{-2}$. 
This solution was used as the starting value for a least squares fit
to the Q and U measurements in each pixel.
The product of this is a cube with the RM of the peak, the EVPA at $\lambda$=0, the peak polarized intensity and error estimates.

\section{Data Limitations\label{Limitation}}
The resultant survey images are limited by a number of effects.
\subsection{Dynamic Range}
Individual pointing images could be limited in dynamic range by
Direction Dependent Gain Effects (DDE) from particularly bright
sources, especially when there were multiple such sources and/or they
were far from the pointing center.
The direction independent self calibration applied to all data is
inadequate to reduce the DDEs so ``Peeling'' \citep{Noordam2004} was
used to improve selected pointings included in the mosaics.
No DDE corrections were used for the single pointing images and these effects
appear in the raw visibility data in all cases.

\subsection{Largest angular scale}
The largest scale structure that can be recovered from interferometry
data is limited by the length and distribution of the shortest
baselines in units of wavelengths.
Thus, the largest recoverable scale is frequency dependent and depends
on the detailed distribution of emission.
Structures no larger than $10'$ are generally well reconstructed but
larger scale structure may not be fully represented.
Since the apparent loss of extended emission is frequency dependent,
the derived spectral index\footnote{The spectral index, $\alpha$, is defined as $I_\nu \propto \nu^{\alpha}$.} of extended emission may be MUCH (2 or 3 in
$\alpha$) steeper than the true value.

\subsection{Astrometry}
Systematic astrometric errors are largely due to the low accuracy
geometric model used in the correlator. 
Phases are not accurately transferred over the large distances needed
to get to a suitably strong calibrator and systematic offsets can be
up to the order of 1\asec. 
The mosaics have been corrected for the most serious errors but the
individual pointing images and the raw visibility data have not.

\subsection{Flux Density and Spectral Index}
The limited short baseline coverage limits the amount of flux density
than can be recovered from extended regions of emission; the flux
densities of these extended features will be underestimated.
This effect is frequency dependent as it depends on the
baseline lengths in wavelengths which varies by approximately a factor
of two across the MeerKAT bandpass.
As the amount of recovered flux density from significantly resolved
structures decreases with increasing frequency, the result is possibly
a VERY significant steepening of the apparent spectral index.
There is limited emission on scales large enough for this effect to be
important in the images presented.

\subsection{Array Primary Beamshape}
The effective primary beam shape of the array is not the same as that
of a given antenna as pointing errors are sufficiently large to
broaden the array beam. 
This causes errors in apparent flux density in the single pointing
data far from the pointing position where the errors in the assumed
array beamshape can also cause unphysical effects in the apparent
spectrum. 
Except near the edges of the mosaics, the flux densities are dominated by pointings
close to any given position; this reduces the errors from an
inadequate primary beam correction.

\subsection{Polarization}
The MeerKAT beam has significant off-axis and higher frequency
instrumental polarization.
To some degree this is reduced in the mosaics which are heavily
weighted toward the beam center and a given position on the sky is
viewed through various parts of the beam in the different overlapping 
pointings. 
In the single pointings the less reliable portions of the beam are
blanked out in the Faraday rotation images but may still be a problem
at the higher frequencies.
These effects are uncorrected in the raw visibility data.

\section{Data Products\label{Products}}
The raw data is available from the SARAO archive
(https://archive.sarao.ac.za/) under project code SSV-20180505-FC-01.
Image products (gzipped FITS) and catalogs (csv) can be obtained from \url{https://doi.org/10.48479/f2a2-qw16}.

\subsection{Mosaics}
 The mosaics are provided in full cube and 5 plane ``FitSpec" versions.
The latter have  planes: 1) broadband flux density, 2) spectral index
where fitted else blanked, 3) least  squares error for flux density,
4) least squares  error for the fitted spectral index, 5) $\chi^2$ per
degree of freedom.  
The mosaic and single pointing full cubes are in the format described
by \cite{MFImage}.
The full cube and FitSpec versions have the same broadband and
spectral index planes which are blanked outside of the area covered by
highest frequency bin. 
The frequencies of the planes in the mosaic sub--band cube
images are given in Table \ref{tab:subband} and the effective
frequency of the broadband images is 1333.13 MHz.

The first plane of the full cube images is the broadband image, second
plane is the spectral index (Stokes I only) and subsequent planes are
the sub--band images. 
The names of the full cube images are 
G$<$lll.l$><\pm$bb.b$>$\_$<$s$>$.fits.gz where $<$s$>$ is the Stokes
parameter (``I'',``Q'', ``U'', or ``V''), $<$lll.l$>$ is the central
Galactic longitude, and $<\pm$bb.b$>$ the central latitude. 
Images with the parameters from the Stokes I least squares spectral
fit (``FitSpec'') have names G$<$lll.l$><\pm$bb.b$>$\_FitSpec.fits.gz
 
The peak Faraday depth (Rotation Measure=RM) images were
derived from a pixel-by-pixel search in RM space $\pm$150 rad/m$^2$
with an increment of 0.5 rad/m$^2$ which is used as the starting
solution in a nonlinear least squares fit.
The planes are  1) RM (rad m$^{-2}$), 2) EVPA@$\lambda$=0 (rad), 3)
Polarized intensity (Jy), 4) least squares error of RM, 5) least
squares error of EVPA, 6) $\chi^2$ per degree of freedom. 
Names are of the form G$<$lll.l$><\pm$bb.b$>$\_RM.fits.gz.
The first two planes (as well as 4 and 5) are blanked where the
polarized intensity (3rd plane) is below 40 $\mu$Jy beam$^{-1}$.

\begin{table}[h]
    \centering
    \begin{tabular}{c|c|c}
    Subband & Frequency & Comment \\
             & MHz      & \\
             \hline
       0 &  1333.13& Broadband\\
       2 &  908.0 & \\
       3 &  952.3 & \\
       4 &  996.6 & \\
       5 & 1043.4 & \\
       6 & 1092.8 & \\
       7 & 1144.6 & \\
       8 & 1198.9 & Blanked\\
       9 & 1255.8 & Blanked\\
       10 & 1317.2 & \\
       11 & 1381.2 & \\
       12 & 1448.1 & \\
       13 & 1519.9 & \\
       14 & 1593.9 & \\
       15 & 1656.2 & \\
       \hline
    \end{tabular}
    \caption{MeerKAT sub-band central frequencies.}
    \label{tab:subband}
\end{table}

\subsection{Single Pointing Sample}
The single pointing sample images have full field primary beam
corrected Stokes I ``Cubes'' with names of the form
G$<$lll$\pm$bb$>$\_I\_PB.fits.gz where 
lll and $\pm$bb are the Galactic longitude and latitude of the
pointing center.  
The Stokes Q, U, and V images, without primary beam corrections, are in
files with names of the form
G$<$lll$\pm$bb$>$\_$<$s$>$.fits.gz where s is ``Q", ``U", or ``V".
The coordinates are Equatorial and these images have the same
channelization as the mosaics, that given in Table \ref{tab:subband}.
The effective frequency of the broadband images is 1333.13 MHz.

There is also a spectral fit primary beam corrected version
(G$<$lll$\pm$bb$>$\_FitSpec.fits.gz) similar to the ``FitSpec" mosaic images. 
These are limited to the area  within the $\geq$35\% gain of the antenna pattern at 1.333 GHz (radius=41\amin).  The broadband image and spectral index are produced in the same way as used for the mosaics.   
   RM cubes are similar to the RM cubes of the mosaics but are also
limited to the region of $\geq$35\% antenna gain.
They are not primary beam corrected and have names
G$<$lll$\pm$bb$>$\_RM.fits.gz .
The first two planes are blanked where the polarized intensity (3rd
plane) is below 50 $\mu$Jy beam$^{-1}$.

\subsection{Source Catalogs}
Source catalogs of compact sources are provided in csv format.  Separate lists are given for the mosaics and combined single pointing images.  Basic source parameters (position, flux density, size...) are given in one set of listings (Mosaic\_Cat.csv, Single\_Cat.csv) and the spectral index and polarization for sources with detected polarization in another (Mosaic\_Pol\_Cat.csv and Single\_Pol\_Cat.csv).
Samples are given in Tables \ref{table:BasicMosaic}, \ref{table:BasicOff}, \ref{table:PolnMosaic} and \ref{table:PolnOff}.
The list of extended sources examined as possible Giant Radio Galaxies is given in Extended.csv.  A sample is given in Table \ref{table:ExtendedCat}.

\section{Results\label{Results}}
There is little Galactic total intensity emission away from the plane.
The Milky Way is largely transparent at the wavelength of these
observations and the extragalactic sky is visible even near the Galactic plane.
A visual search of the images did not reveal any further features similar to the MeerKAT Bubbles, even when imaged at lower resolution using natural weighting (24\asec\ FWHM).
The noise varies with Galactic location but the typical Stokes I RMS is $\sim$20 $\mu$Jy beam$^{-1}$, $\sim$10 $\mu$Jy beam$^{-1}$ in Stokes Q and U and $\sim$8 $\mu$Jy beam$^{-1}$ in Stokes V.
\subsection{Mosaics\label{Mosaics}}
An analysis of these mosaics searching for compact, polarized, steep
spectrum sources searching for pulsars and other Galactic stars is given in \cite{Frail2024}.
\subsubsection{Compact Source Catalog\label{Catalog}}
The relatively compact sources in the mosaic fields were
cataloged using Obit program FndSou. 
Islands of emission were identified and were fitted with elliptical
Gaussian shaped components.
No background corrections were made.
An error analysis of the fitted components followed that of
\cite{Condon1997}. 
A visual inspection of the results identified entries which were clear
artifacts and they were deleted from the list and missing sources were
added. 
Catalog entries were restricted to have a fitted FWHM of 50\asec\ or less and to those which have a local (within a box of half width 50 pixels) signal-to-noise ratio of at least 5.
The values and error estimates of the other data products (RM, PPol,
VPol, SI) were interpolated at the position of the peak Stokes I in
the appropriate image.
Source names and positions are given in Equatorial (J2000) coordinates
as the vast majority of individual objects are extragalactic and even
for Galactic objects, Equatorial coordinates are preferred for
cross--identification.
No attempt has been made to associate multiple components in the derived catalogs into physical sources.

The basic position and size information of a sample of the component
list is given in Table \ref{table:BasicMosaic} and a sample of the
polarization values for sources with detected polarization in Table \ref{table:PolnMosaic}. 
Sources with a fitted major axis size larger than 15" were excluded from Table \ref{table:PolnMosaic}.  Entries were required to have at least 1\% linear or 2\% circular polarization to minimize the effects of residual instrumental polarization.
The fitting in rotation measure results has a more significant bias in
the measure of the linearly polarized intensity than the simple
derivation from Q and U.
The mode of the distribution of the maximum polarized intensity per
pixel in each of the images was used as the polarization bias for that
image. 
The vast majority of pixels have no detectable polarization.
Duplicate entries from the overlapping regions of the mosaics were
removed. 
The full lists are available in doi \url{https://doi.org/10.48479/f2a2-qw16} as Mosaic\_Cat.csv and Mosaic\_Pol\_Cat.csv.

\begin{longrotatetable}
\begin{deluxetable*}{c|cc|cc|c|c|r|rr|c|c|r}
  \tablecaption{Catalog of Mosaics: Sample of Basic Data}
  \tablehead{
    \colhead{Name\tablenotemark{a}} & \colhead{RA (2000)} & \colhead{$\pm$} & \colhead{Dec (2000)} & \colhead{$\pm$} &
    \colhead{Glong} & \colhead{Glat} & \colhead{Peak\tablenotemark{b}}  & \colhead{Total\tablenotemark{c}} & \colhead{$\pm$} & 
    \colhead{Size\tablenotemark{d}} & \colhead{PA\tablenotemark{e}} & \colhead{Mosaic\tablenotemark{f}}\\
\colhead{ } & \colhead{h m s} & \colhead{s} & \colhead{d m s} & \colhead{\asec} & \colhead{$^\circ$} & \colhead{$^\circ$} & 
 \colhead{$\mu$Jy/bm} & \colhead{} & \colhead{$\mu$Jy/bm} & \colhead{\asec} & 
 \colhead{$^\circ$} & \colhead{}
}
\startdata
 J1635404-183315 &   16 35 40.47 & 0.027 & -18 33 15.0 & 0.35 & 359.230 & 19.026  &   243.3 &   258.1 &  19.8   & $\leq$ 7.1          &        & G000.0+17.5\\
 J1635416-182947 &   16 35 41.62 & 0.030 & -18 29 47.6 & 0.59 & 359.281 & 19.058  &   171.8 &   185.6 &  18.5   & $\leq$ 8.7          &        & G000.0+17.5\\
 J1635419-183727 &   16 35 41.92 & 0.029 & -18 37 27.7 & 0.81 & 359.176 & 18.977  &   182.7 &   278.9 &  34.9 &     9.4 $\times$  $\leq$ 5.2  &   12.5 & G000.0+17.5\\
 J1635437-182934 &   16 35 43.75 & 0.060 & -18 29 34.8 & 0.80 & 359.290 & 19.054  &   125.0 &   144.8 &  19.8   & $\leq$12.8          &        & G000.0+17.5\\
 J1635458-182943 &   16 35 45.83 & 0.006 & -18 29 43.3 & 0.09 & 359.293 & 19.046  &   839.9 &   849.7 &  29.5   & $\leq$ 3.4          &        & G000.0+17.5\\
 J1636106-191811 &   16 36 10.60 & 0.029 & -19 18 11.7 & 0.47 & 358.692 & 18.464  &   223.6 &   293.2 &  30.7 &     6.4 $\times$  $\leq$ 6.0  &  -34.1 & G000.0+17.5\\
 J1636142-191409 &   16 36 14.21 & 0.004 & -19 14 09.3 & 0.06 & 358.756 & 18.496  &  1695.4 &    2245 &    81 &     5.8 $\times$   4.6   &    5.3 & G000.0+17.5\\
 J1639412-175941 &   16 39 41.25 & 0.023 & -17 59 41.2 & 0.29 &   0.318 & 18.625  &   194.8 &   252.3 &  19.4 &     5.7 $\times$  $\leq$ 6.1  &   66.1 & G000.0+17.5\\
 J1639414-195130 &   16 39 41.45 & 0.008 & -19 51 30.5 & 0.12 & 358.779 & 17.471  &   885.6 &   982.8 &  45.3 &     3.9 $\times$  $\leq$ 3.0  &  -46.5 & G000.0+17.5\\
 J1646082-193755 &   16 46 08.23 & 0.021 & -19 37 55.1 & 0.28 & 359.942 & 16.416  &   189.9 &   196.5 &  12.8   & $\leq$ 5.6          &        & G000.0+17.5\\
 J1647572-185957 &   16 47 57.23 & 0.014 & -18 59 57.0 & 0.19 &   0.737 & 16.459  &   359.0 &   408.3 &  24.3 &     4.3 $\times$  $\leq$ 3.7  &  -49.5 & G000.0+17.5\\
 J1647574-200113 &   16 47 57.41 & 0.070 & -20 01 13.8 & 1.14 & 359.892 & 15.840  &   110.7 &   241.4 &  33.7 &    15.3 $\times$  $\leq$ 6.3  &  -40.7 & G000.0+14.5\\
 J1718203-252734 &   17 18 20.31 & 0.030 & -25 27 34.8 & 0.36 & 359.610 &  6.979  &   209.8 &   234.7 &  18.8   & $\leq$ 8.1          &        & G000.0+5.5\\
  \enddata
\tablenotetext{a}{Source names are derived from the integral parts of the hours, minutes and seconds and the tenth of seconds of RA with the declination given by the integral parts of degrees, minutes and seconds of declination.}
\tablenotetext{b}{The peak flux density.}
\tablenotetext{c}{The integrated flux density.}
\tablenotetext{d}{Deconvolved angular size or upper limit.}
\tablenotetext{e}{Position angle of major axis of deconvolved ellipse.}
\tablenotetext{f}{Name of mosaic field containing source.}
\label{table:BasicMosaic}
\end{deluxetable*}
\end{longrotatetable}

\begin{longrotatetable}
\begin{deluxetable*}{|c|rr|rr|rr|rr|rr|c|r|r|}
  \tablecaption{Catalog of Mosaics: Sample of Polarization Data}
  \tablehead{
    \colhead{Name} & \colhead{IPol\tablenotemark{a}} & \colhead{$\pm$} & \colhead{SI\tablenotemark{b}} & \colhead{$\pm$} &\colhead{PPol\tablenotemark{c}} & \colhead{$\pm$} &
    \colhead{RM\tablenotemark{d}} & \colhead{$\pm$} & \colhead{VPol\tablenotemark{e}} & \colhead{$\pm$} & 
    \colhead{res\tablenotemark{f}} & \colhead{Fract P\tablenotemark{g}} & \colhead{Fract V\tablenotemark{h}} \\
\colhead{ } & \colhead{$\mu$Jy/bm} & \colhead{} & \colhead{} &\colhead{} &  \colhead{$\mu$Jy/bm} & \colhead{} & \colhead{rad m$^{-2}$} & 
\colhead{} & \colhead{$\mu$Jy/bm} & \colhead{} & \colhead{} & \colhead{\%} &  \colhead{\%}
}
\startdata
 J1636142-191409 &     2245 &     81  &  -1.116 & 0.040 &  137.4 &  10.3&    -18.0 &   1.1&\phantom{1.0}&\phantom{1.0}&  Y  &        8.1&\phantom{1.0} \\
 J1639412-175941 &    252.3 &   19.4  &   \phantom{1.000} & \phantom{1.000}  &   45.7 &   9.0&    -14.9 &   2.3&\phantom{1.0}&\phantom{1.0}&  Y  &       23.5&\phantom{1.0} \\
 J1639414-195130 &    982.8 &   45.3  &  -0.330 & 0.081 &  131.5 &  10.1&     -7.8 &   1.1&\phantom{1.0}&\phantom{1.0}&  Y  &       14.8&\phantom{1.0} \\
 J1640436-195948 &    440.1 &   37.9  &  -0.968 & 0.222 &   77.8 &  10.8&     -5.1 &   1.6&\phantom{1.0}&\phantom{1.0}&  Y  &       25.2&\phantom{1.0} \\
 J1653366-212216 &    150.9 &   11.9  &   \phantom{1.000} & \phantom{1.000}  &   81.1 &  14.4&    -10.3 &   7.1&\phantom{1.0}&\phantom{1.0}&  N  &       58.6&\phantom{1.0} \\
 J1718203-252734 &    234.7 &   18.8  &  -0.639 & 0.397 &\phantom{1.0}&\phantom{1.0}&\phantom{1.0} & \phantom{1.0} &  -30.5 &   7.2&  N  &\phantom{1.0}&      -14.6 \\
 J1718473-263842 &     1979 &     69  &  -1.030 & 0.059 &  220.2 &  16.2&   -114.8 &   1.0&\phantom{1.0}&\phantom{1.0}&  Y  &       18.9&\phantom{1.0} \\
 J1719566-245703 &    284.9 &   17.4  &   1.721 & 0.320 &\phantom{1.0}&\phantom{1.0}&     76.7 &   4.7&  -82.9 &   7.6&  N  &\phantom{1.0}&      -29.1 \\
 J1720341-293315 &     2795 &     85  &  -1.050 & 0.032 &  225.6 &  16.0&      8.5 &   0.9& -160.2 &   8.6&  N  &        8.1&       -5.7 \\
 J1720410-244354 &    896.5 &   32.0  &  -0.693 & 0.079 &  122.2 &  13.7&   -139.3 &   1.7&\phantom{1.0}&\phantom{1.0}&  N  &       13.8&\phantom{1.0} \\
 J1720550-272040 &    287.4 &   17.1  &  -0.009 & 0.348 &\phantom{1.0}&\phantom{1.0}&     62.1 &   5.2&  -85.7 &   7.0&  N  &\phantom{1.0}&      -31.9 \\
 J1721481-242511 &    20353 &    611  &  -0.795 & 0.003 & 2794.3 &  16.5&    -67.3 &   0.1&\phantom{1.0}&\phantom{1.0}&  N  &       14.0&\phantom{1.0} \\
   \enddata
\tablenotetext{a}{Integrated flux density.}
\tablenotetext{b}{Spectral index.}
\tablenotetext{c}{Linearly polarized peak intensity.}
\tablenotetext{d}{Faraday rotation measure.}
\tablenotetext{e}{Circularly polarized peak intensity.}
\tablenotetext{f}{Is the source resolved? ``Y'' or ``N''}
\tablenotetext{g}{Fractional linear polarization, min (P/I)=1\%.}
\tablenotetext{h}{Fractional circular polarization, min($|$V$|$)/I=2\%.}
\label{table:PolnMosaic}
\end{deluxetable*}
\end{longrotatetable}

\subsubsection{Faraday Screen}
\begin{figure}
\includegraphics[width=2.5in]{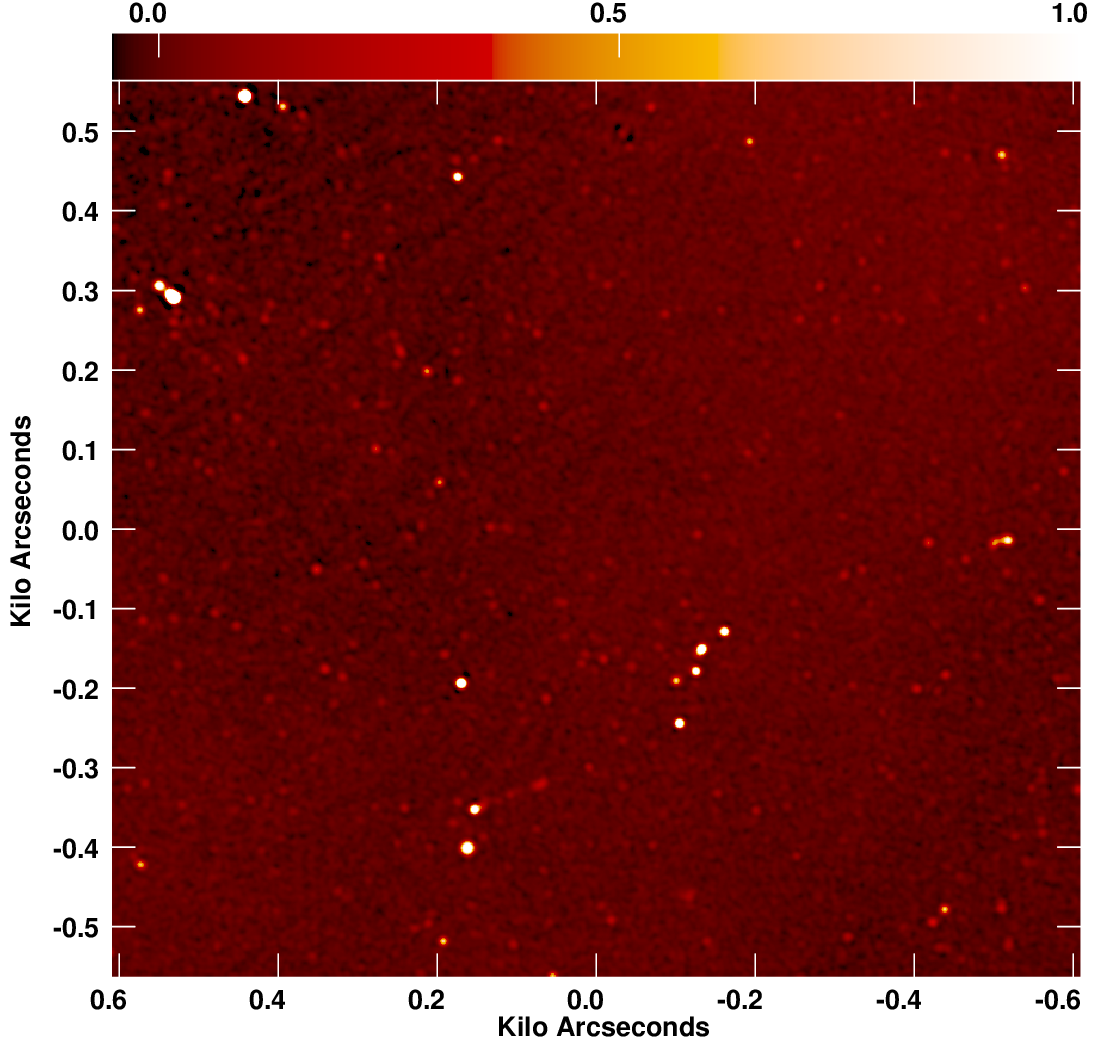}
\includegraphics[width=2.5in]{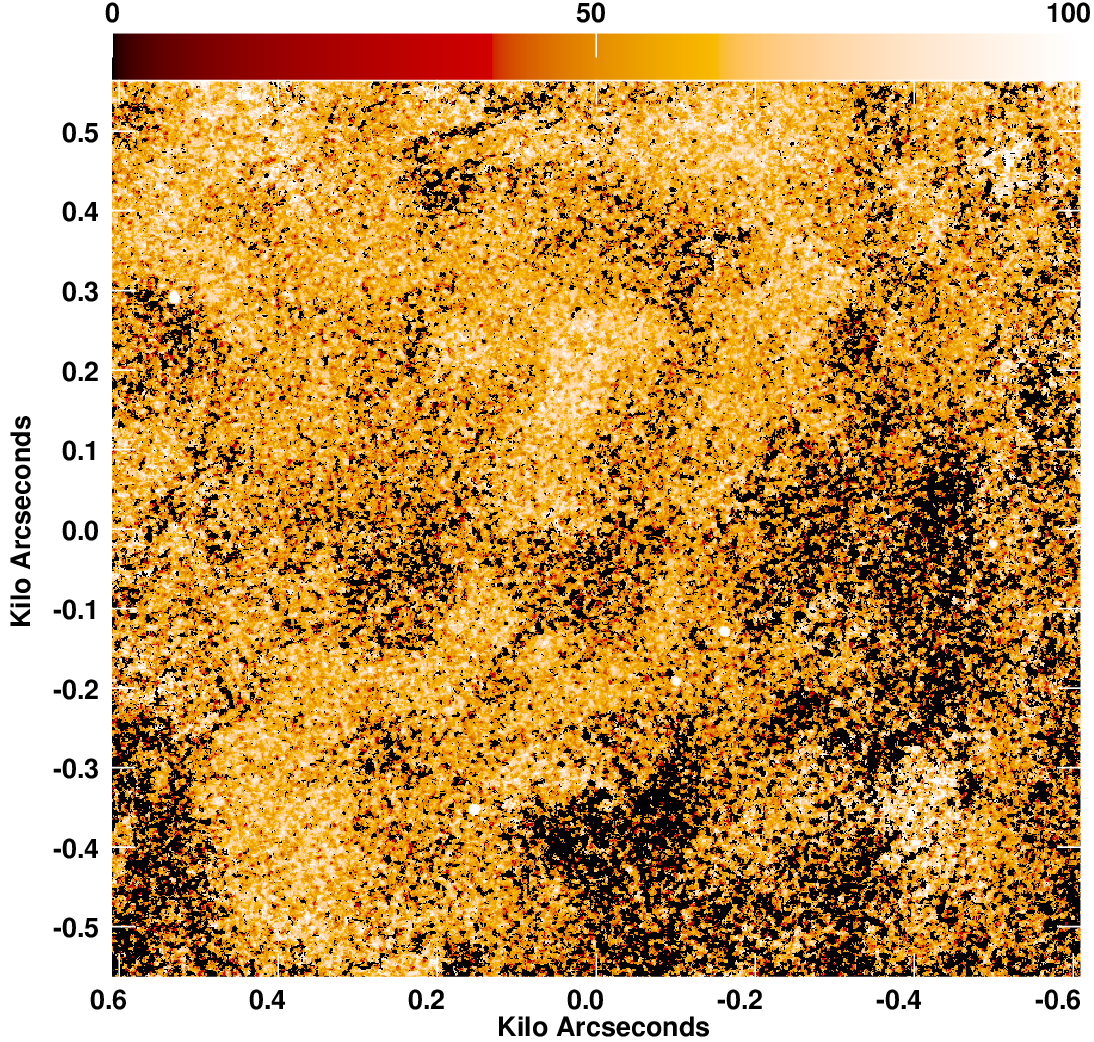}
\includegraphics[width=2.5in]{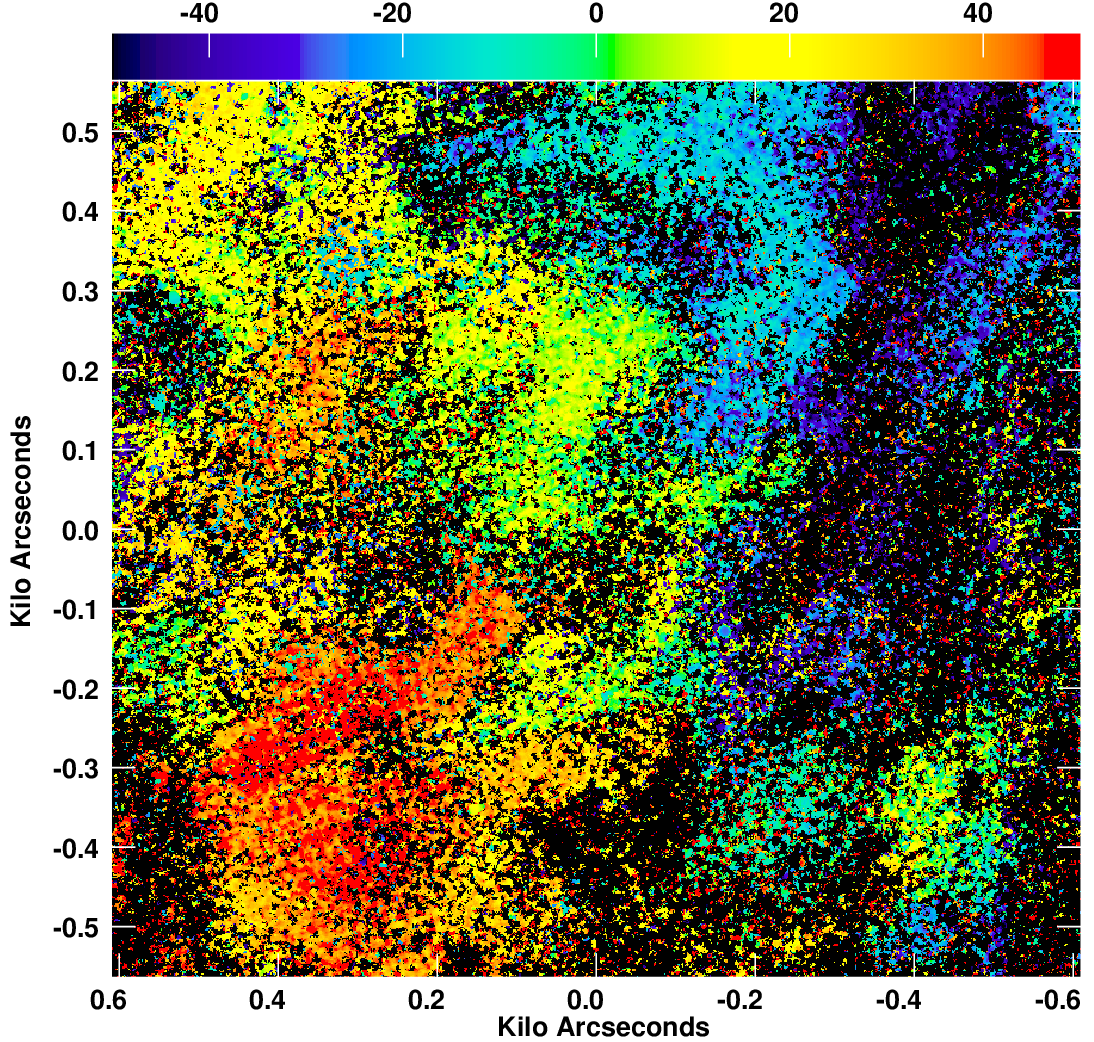}
\caption{ The region around G003.55+5.38 showing the Faraday screen
  in the ISM in front of the smooth background Galactic polarized disk
  emission. 
The region shown is 9\amin $\times$ 8.\amin 3 in extent.\\
{\bf Top:} Stokes I, scalebar at top labeled in mJy beam$^{-1}$.
{\bf Middle:} Polarized intensity, scalebar at top labeled in $\mu$Jy beam$^{-1}$.
{\bf Bottom:} Rotation measure, scalebar at top labeled in radians m$^{-2}$.
}
\label{fig:FaradayScreen}
\end{figure}

There is a Faraday screen in the ISM that has structure on a fine enough scale that Stokes Q and U are not filtered out in the way that the much smoother total intensity emission is.
One example of this is shown in Figure \ref{fig:FaradayScreen}.
The bulk of the polarized emission is unrelated to any feature seen in
total intensity.

\subsubsection{Particularly Extended Sources}
Some of the sources in these fields are sufficiently extended to warrant individual discussion.
The most extended of these, J1712427-243550, will be covered in a separate publication. 
See Section \ref{GRGs} for a discussion of the determination of redshifts and source sizes of these and other sources.
These are background AGNs.
\par
\begin{figure}
\includegraphics[width=3.5in]{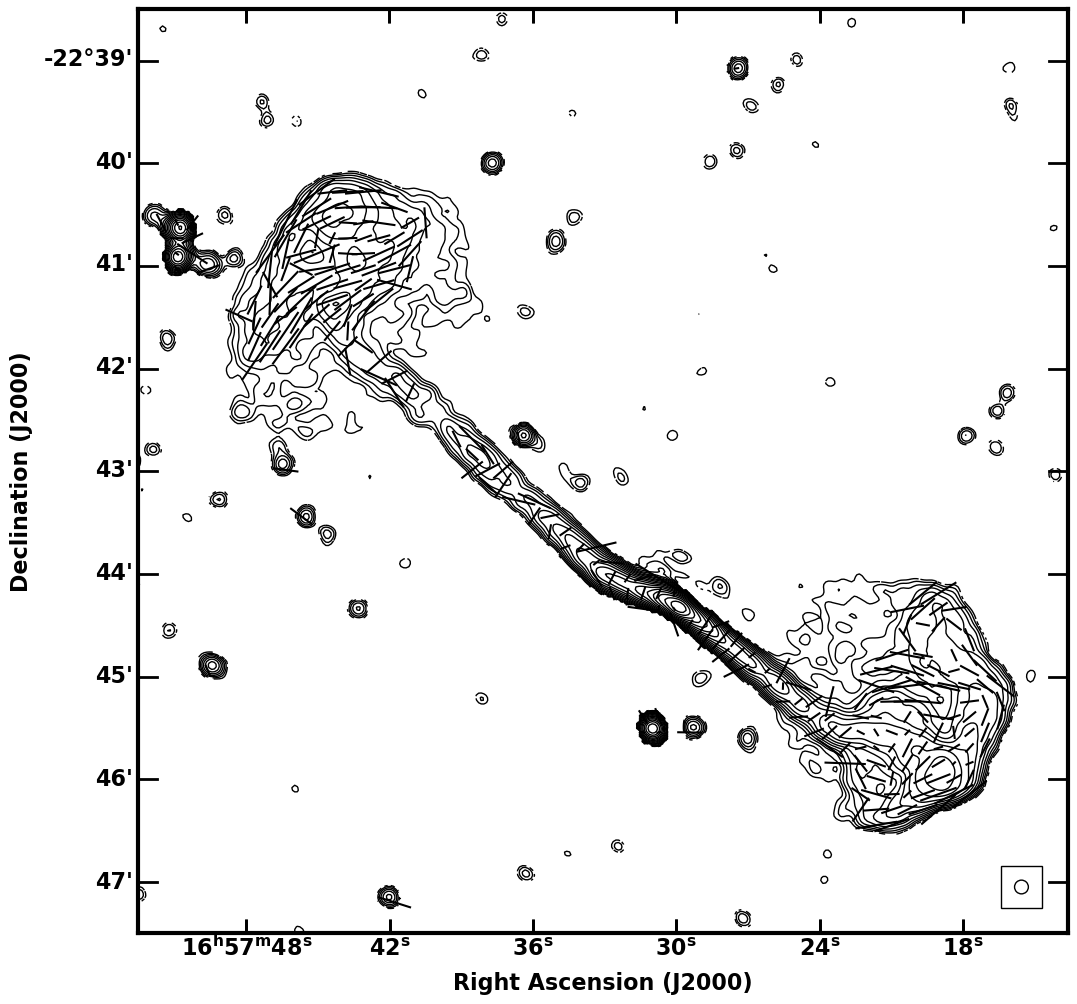}
\caption{J1657300-224420: Contours are at powers of $\sqrt{2}$ from 36.6\,$\mu$Jy/bm \chga{(lowest, outermost contour)}. The vector lengths are proportional to the fractional polarization, shown as ``B'' vectors. The restoring beam FWHM is given in the box in the lower right corner.}
\label{fig:J1657-2244}
\end{figure}

\noindent
{\bf J1657300-224420:} 
The source shown in Figure \ref{fig:J1657-2244} is an FRII AGN, characterized by its prominent double-lobed structure, with the core located at RA(2000) = 16 57 30.02, Dec = -22 44 20.4. The radio lobes extend along a northeast-southwest axis, culminating in bright hotspots at their edges. Contours delineate regions of intense radio emission, while the polarization ``B'' vectors overlaying the image provide insights into the magnetic field alignment across the lobes.
The total flux density at 1333.1 MHz is 0.211 Jy, the angular extent of the source is 4.\amin26 and the redshift (NED) is 0.035164.
\par
\begin{figure}
\includegraphics[width=3.5in]{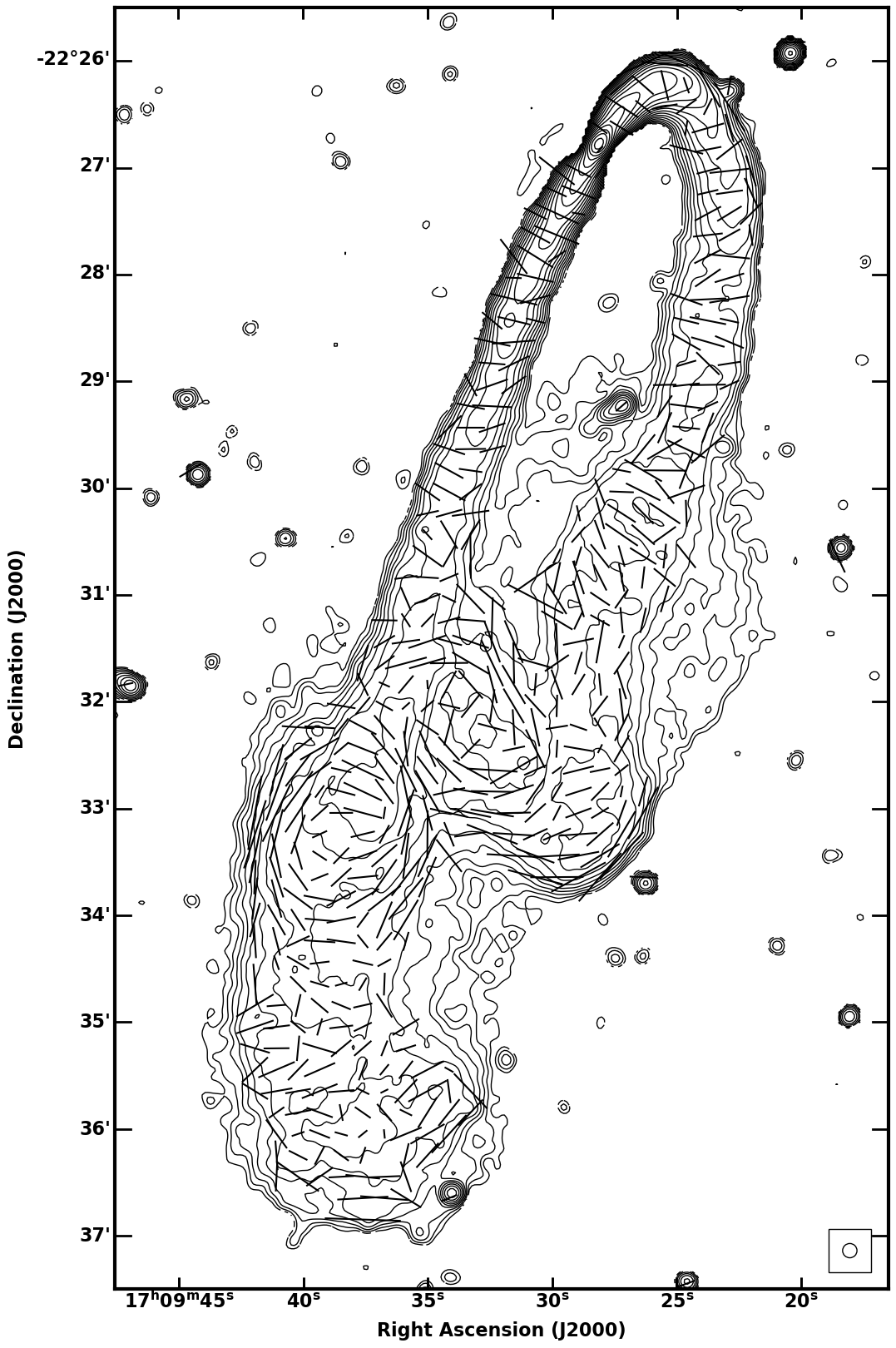}
\caption{J1709281-222649: Contours are at powers of $\sqrt{2}$ from 35\,$\mu$Jy/bm \chga{(lowest, outermost contour)}. The vector lengths are proportional to the fractional polarization, shown as ``B'' vectors. The restoring beam FWHM is given in the box in the lower right corner.}
\label{fig:J1709-2226}
\end{figure}

\noindent
{\bf J1709281-222649:}
The source shown in Figure \ref{fig:J1709-2226} is a narrow-angle tail radio galaxy, with a compact core located at RA(2000) = 17 09 28.15, Dec = -22 26 49.8. Its distinct double-plumed morphology displays elongated jets stretching predominantly toward the south. The superposed polarization ``B'' vectors trace the magnetic field distribution across the extended structure, while the contours depict variations in radio brightness across the source.
The integrated flux density at 1333.1 MHz is 0.636 Jy, the angular extent following bends is 19.\amin48 and the redshift (NED) is 0.03955.
\par
\begin{figure}
\includegraphics[width=3.5in]{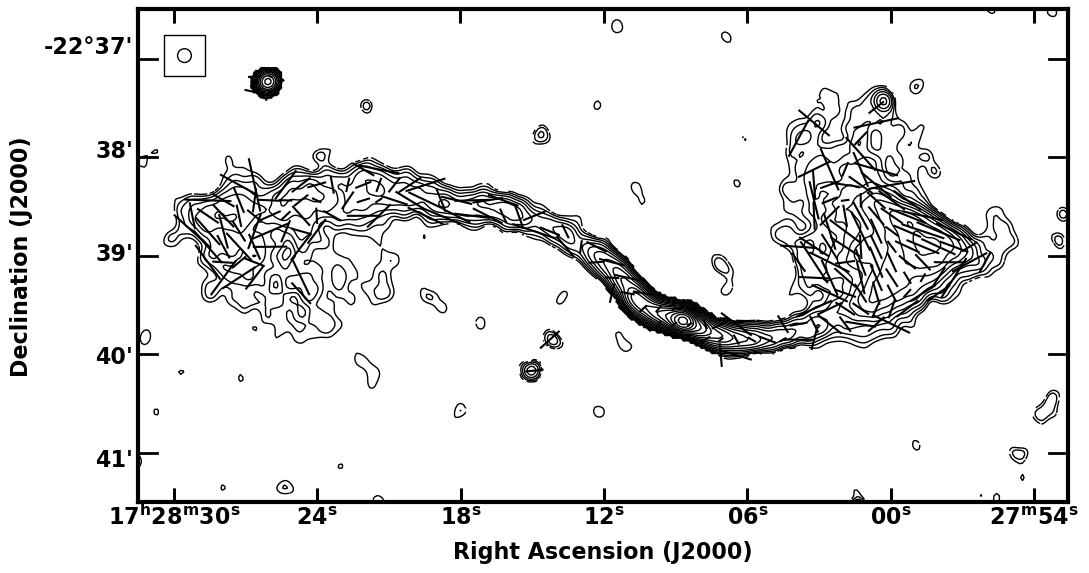}
\caption{J1728088-223940: Contours are at powers of $\sqrt{2}$ from 63\,$\mu$Jy/bm \chga{(lowest, outermost contour)}. The vector lengths are proportional to the fractional polarization, shown as ``B'' vectors. The restoring beam FWHM is given in the box in the top left corner.}
\label{fig:J1728-2239}
\end{figure}

\noindent
{\bf J1728088-223940:}
The source shown in Figure \ref{fig:J1728-2239} is an FRI AGN, featuring a complex east-west oriented double-plume structure, anchored by a core located at RA(2000) = 17 28 08.76, Dec = -22 39 40.4. The plumes display noticeable asymmetry, with the western plume appearing more spread out than the eastern one, suggesting possible environmental differences on either side. Contour levels highlight the intensity gradients, while polarization ``B'' vectors provide critical insights into the magnetic field configurations across the jets and plumes.
The total flux density at 1333.1 MHz is 0.225 Jy and the angular extent following bends is 9.\amin31.
\par
\begin{figure}
\includegraphics[width=3.5in]{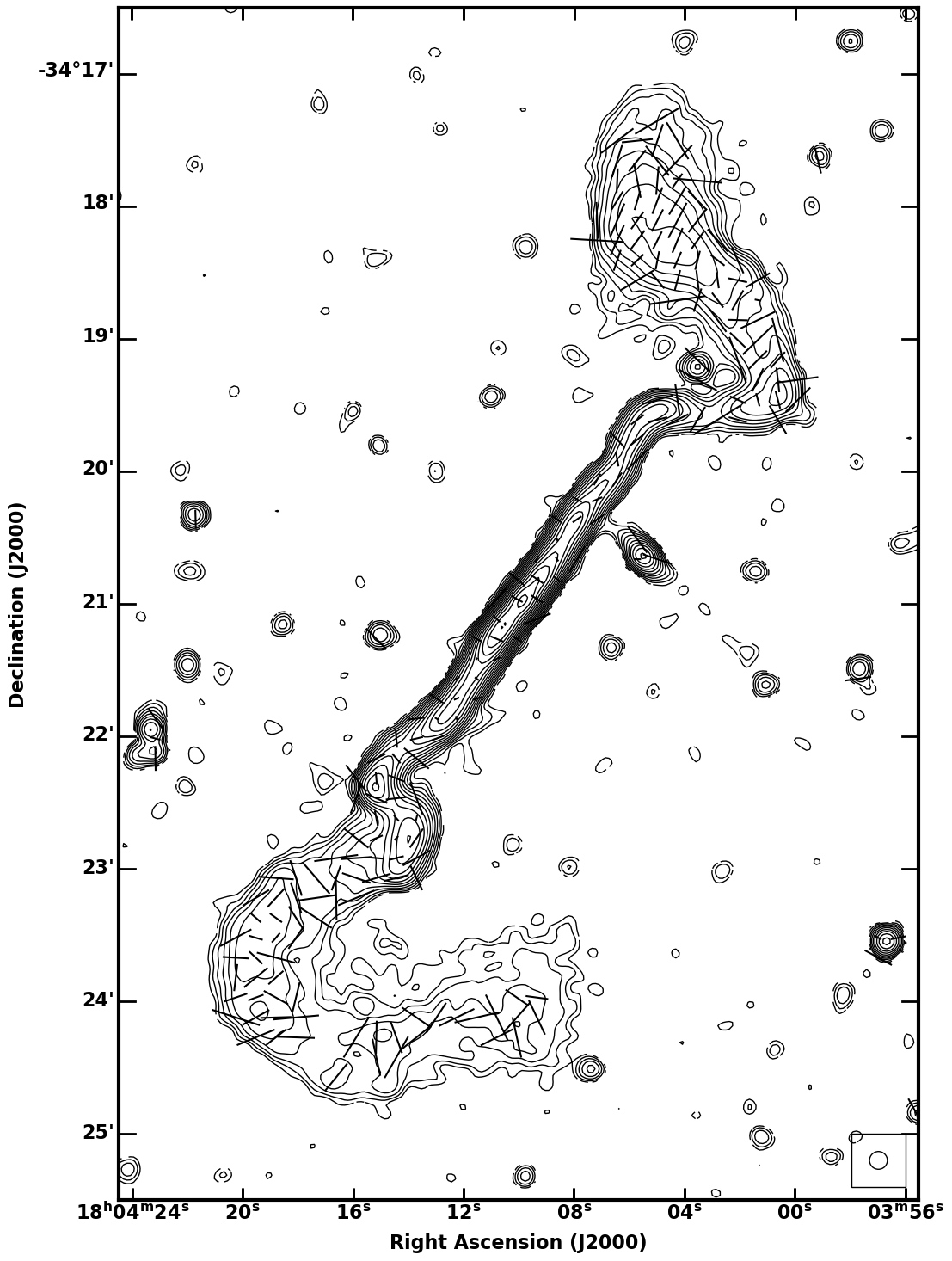}
\caption{J1804104-342107: Contours are at powers of $\sqrt{2}$ from 27\,$\mu$Jy/bm \chga{(lowest, outermost contour)}. The vector lengths are proportional to the fractional polarization, shown as ``B'' vectors. The restoring beam FWHM is given in the box in the lower right corner.}
\label{fig:J1804-3421}
\end{figure}

\noindent
{\bf J1804104-342107:}
The source shown in Figure \ref{fig:J1804-3421} is an FRI AGN, displaying a striking northwest-southeast bent-lobed morphology. The core, located at RA(2000) = 18 04 10.37, Dec = -34 21 07.3, anchors the lobes, which terminate in pronounced hotspots. The noticeable bends in the hotspots at the jet termini suggest potential interactions with the surrounding environment. Contours map the radio intensity distribution, while the polarization ``B'' vectors highlight the magnetic field orientation throughout the structure.
The integrated flux density of the source is 0.125 Jy and the angular extent following bends is 12.\amin88.

\subsection{Single Pointing Samples}

\subsubsection{Compact Source Catalog}
The relatively compact sources in the single pointing fields were
cataloged as for the mosaics, see Section \ref{Catalog}. 
The basic position and size information of a sample from the combined
lists is given in Table \ref{table:BasicOff} and sample of the
polarization values in Table \ref{table:PolnOff}. 
Sources with a fitted major axis size larger than 15\asec\ were excluded from the polarization listing, Table \ref{table:PolnOff}.
Entries in Table \ref{table:PolnOff} were required to have either a linear polarization in excess of 1\% or a circular polarization in excess of 2\% to minimize the effects of residual instrumental polarization.
The full lists are available in doi \url{https://doi.org/10.48479/f2a2-qw16} as Single\_Cat.csv and Single\_Pol\_Cat.csv.
\begin{longrotatetable}
\begin{deluxetable*}{c|cc|cc|r|r|rr|c|c|c}
  \tablecaption{Catalog of Single Pointings: Sample of Basic Data}
  \tablehead{
    \colhead{Name\tablenotemark{a}} & \colhead{RA (2000)} & \colhead{$\pm$} & \colhead{Dec (2000)} & \colhead{$\pm$} &
    \colhead{Gain\tablenotemark{b}} & \colhead{Peak\tablenotemark{c}} & \colhead{Total\tablenotemark{d}} & \colhead{$\pm$} & 
    \colhead{Size\tablenotemark{e}} & \colhead{PA\tablenotemark{f}} & \colhead{Field\tablenotemark{g}}\\
\colhead{ } & \colhead{h m s} & \colhead{s} & \colhead{d m s} & \colhead{\asec} & \colhead{} & \colhead{$\mu$Jy/bm} & \colhead{} & \colhead{$\mu$Jy/bm} & \colhead{\asec} & \colhead{$\circ$} & \colhead{}\\
}
\startdata
 J1349327-412405 &   13 49 32.775 & 0.017 & -41 24 05.51 & 0.22 &  0.55 &   443.9 &   464.3 &  25.0   & $\leq$ 4.8          &        & G15+20\\
 J1349341-412229 &   13 49 34.141 & 0.069 & -41 22 29.50 & 1.18 &  0.55 &   159.2 &   290.8 &  52.1 &    10.1 $\times$  $\leq$ 9.6  &   15.6 & G15+20\\
 J1349345-411858 &   13 49 34.591 & 0.005 & -41 18 58.69 & 0.06 &  0.54 &  1654.4 &    1676 &    55   & $\leq$ 2.8          &        & G15+20\\
 J1349504-412405 &   13 49 50.463 & 0.005 & -41 24 05.92 & 0.06 &  0.63 &  1668.6 &    1699 &    55   & $\leq$ 3.0          &        & G15+20\\
 J1350066-413833 &   13 50 06.685 & 0.049 & -41 38 33.11 & 0.74 &  0.63 &   206.3 &   362.8 &  45.0 &     9.7 $\times$  $\leq$ 7.8  &   27.2 & G15+20\\
 J1350067-414054 &   13 50 06.722 & 0.044 & -41 40 54.16 & 0.63 &  0.61 &   179.2 &   202.7 &  22.6   & $\leq$ 9.5          &        & G15+20\\
 J1402005-462318 &   14 02 00.580 & 0.002 & -46 23 18.48 & 0.02 &  0.60 &  4270.9 &    4307 &   131   & $\leq$ 2.5          &        & G15+15\\
 J1402012-460303 &   14 02 01.217 & 0.089 & -46 03 03.40 & 0.86 &  0.54 &   115.9 &   120.2 &  23.0   & $\leq$ 9.5          &        & G15+15\\
 J1402013-462921 &   14 02 01.329 & 0.001 & -46 29 21.68 & 0.01 &  0.55 &   35536 &   42976 &  1478 &     5.8 $\times$  $\leq$ 2.3  &  -78.3 & G15+15\\
 J1405309-505600 &   14 05 30.911 & 0.041 & -50 56 00.50 & 0.28 &  0.68 &   522.9 &    1115 &    60 &    14.9 $\times$  $\leq$ 3.4  &   56.0 & G15+10\\
 J1405309-504808 &   14 05 30.959 & 0.008 & -50 48 08.50 & 0.08 &  0.60 &  1090.7 &    1120 &    38   & $\leq$ 3.7          &        & G15+10\\
 J1415180-555053 &   14 15 18.096 & 0.010 & -55 50 53.50 & 0.08 &  0.63 &  1349.5 &    1871 &    72 &     7.3 $\times$   2.9   &   47.5 & G15+5\\
 J1845009-140905 &   18 45 00.902 & 0.005 & -14 09 05.94 & 0.08 &  0.67 &  1265.7 &    1430 &    58 &     3.7 $\times$  $\leq$ 3.3  &  -45.2 & G20-5\\
 J2023032+005824 &   20 23 03.270 & 0.034 &  00 58 24.81 & 0.72 &  0.79 &   166.8 &   181.3 &  22.3   & $\leq$ 9.5          &        & G45-20\\
  \enddata
\tablenotetext{a}{Source names are derived from the integral parts of the hours, minutes and seconds and the tenth of seconds of RA with the declination given by the integral parts of degrees, minutes and seconds of declination.}
\tablenotetext{b}{Gain, is the antenna gain, relative to 1 on axis, at the location of the source and has already been applied.}
\tablenotetext{c}{Peak flux density.}
\tablenotetext{d}{Integrated flux density.}
\tablenotetext{e}{Deconvolved major and minor axis sizes or upper limit.}
\tablenotetext{f}{Position angle of the deconvolved major axis.}
\tablenotetext{g}{Name of pointing in which the source appears.}
\label{table:BasicOff}
\end{deluxetable*}
\end{longrotatetable}

\begin{longrotatetable}
\begin{deluxetable*}{|c|rr|rr|rr|rr|rr|c|r|r|}
  \tablecaption{Catalog of Single Pointings: Sample of Polarization Data}
  \tablehead{
    \colhead{Name} & \colhead{IPol\tablenotemark{a}} & \colhead{$\pm$} & \colhead{SI\tablenotemark{b}} & \colhead{$\pm$} &\colhead{PPol\tablenotemark{c}} & \colhead{$\pm$} &
    \colhead{RM\tablenotemark{d}} & \colhead{$\pm$} & \colhead{VPol\tablenotemark{e}} & \colhead{$\pm$} & 
    \colhead{res\tablenotemark{f}} & \colhead{Fract P\tablenotemark{g}} & \colhead{Fract V\tablenotemark{h}} \\
\colhead{ } & \colhead{$\mu$Jy/bm} & \colhead{} & \colhead{} &\colhead{} &  \colhead{$\mu$Jy/bm} & \colhead{} & \colhead{rad m$^{-2}$} & 
 \colhead{} & \colhead{$\mu$Jy/bm} & \colhead{} & \colhead{} & \colhead{} &  \colhead{}
}
\startdata
 J1349504-412405 &     1699 &     55  &  -0.368 & 0.034 &   103.1 &  13.9&    -24.4 &   1.8&\phantom{1.0}&\phantom{1.0}&  N  &        6.2&\phantom{1.0} \\
 J1351500-414240 &    10259 &    352  &  -1.026 & 0.006 &   759.4 &  11.7&    -44.5 &   0.3&\phantom{1.0}&\phantom{1.0}&  N  &        8.7&\phantom{1.0} \\
 J1536100-304116 &    136.3 &   11.9  &   \phantom{1.000} & \phantom{1.000}  &    55.5 &   9.5&     48.5 &   4.6&\phantom{1.0}&\phantom{1.0}&  N  &       43.7&\phantom{1.0} \\
 J1628134-415450 &    171.1 &   23.6  &   \phantom{1.000} & \phantom{1.000}  &    93.5 &  15.5&    -16.3 &   4.0&\phantom{1.0}&\phantom{1.0}&  N  &       61.6&\phantom{1.0} \\
 J1628226-421420 &    381.4 &   27.0  &  -0.567 & 0.180 &   123.8 &  21.9&     32.6 &   6.0&\phantom{1.0}&\phantom{1.0}&  N  &       33.2&\phantom{1.0} \\
 J1719493-020803 &    171.6 &   31.8  &   \phantom{1.000} & \phantom{1.000}  &    65.8 &  11.3&     46.3 &   6.6&\phantom{1.0}&\phantom{1.0}&  N  &       44.1&\phantom{1.0} \\
 J1720572-021223 &     1463 &     49  &  -2.770 & 0.065 &   211.6 &  11.8&      6.2 &   0.9&  121.9 &   9.9&  N  &       14.6&        8.4 \\
 J1838467+152325 &    561.9 &   56.9  &   \phantom{1.000} & \phantom{1.000}  & \phantom{1.0}&\phantom{1.0}&   -138.3 &   3.4&   79.0 &  11.2&  Y  &\phantom{1.0}&       21.5 \\
 J1845009-140905 &     1430 &     58  &  -1.755 & 0.048 & \phantom{1.0}&\phantom{1.0}&    -12.9 &   4.3&  967.2 &  14.8&  Y  &\phantom{1.0}&       76.4 \\
 J1930323+082925 &    123.5 &   23.4  &   \phantom{1.000} & \phantom{1.000}  &    63.3 &  11.8&    -66.8 &   8.7&\phantom{1.0}&\phantom{1.0}&  N  &       55.6&\phantom{1.0} \\
 J1950314+060501 &    428.6 &   22.9  &  -0.691 & 0.141 &    67.4 &  10.5&   -130.6 &   1.6&\phantom{1.0}&\phantom{1.0}&  N  &       16.7&\phantom{1.0} \\
 J2023032+005824 &    181.3 &   22.3  &   \phantom{1.000} & \phantom{1.000}  & \phantom{1.0}&\phantom{1.0}&     -2.2 &   1.9&   99.0 &   9.7&  N  &\phantom{1.0}&       59.3 \\
 \enddata
 \tablenotetext{a}{Integrated intensity.}
 \tablenotetext{b}{Spectral index.}
 \tablenotetext{c}{Linearly polarized peak intensity.}
 \tablenotetext{d}{Faraday rotation measure.}
 \tablenotetext{e}{Circularly polarized peak intensity.}
\tablenotetext{f}{Is the source resolved? ``Y'' or ``N''}
\tablenotetext{g}{Fractional linear polarization, min (P/I)=2\%.}
\tablenotetext{h}{Fractional circular polarization, min($|$V$|$)/I=2\%.}
\label{table:PolnOff}
\end{deluxetable*}
\end{longrotatetable}

\subsubsection{Particularly Extended Sources}
Some of the sources in these fields are sufficiently extended to warrant individual discussion.
These are background AGNs.
\par
\begin{figure}
\includegraphics[width=3.25in]{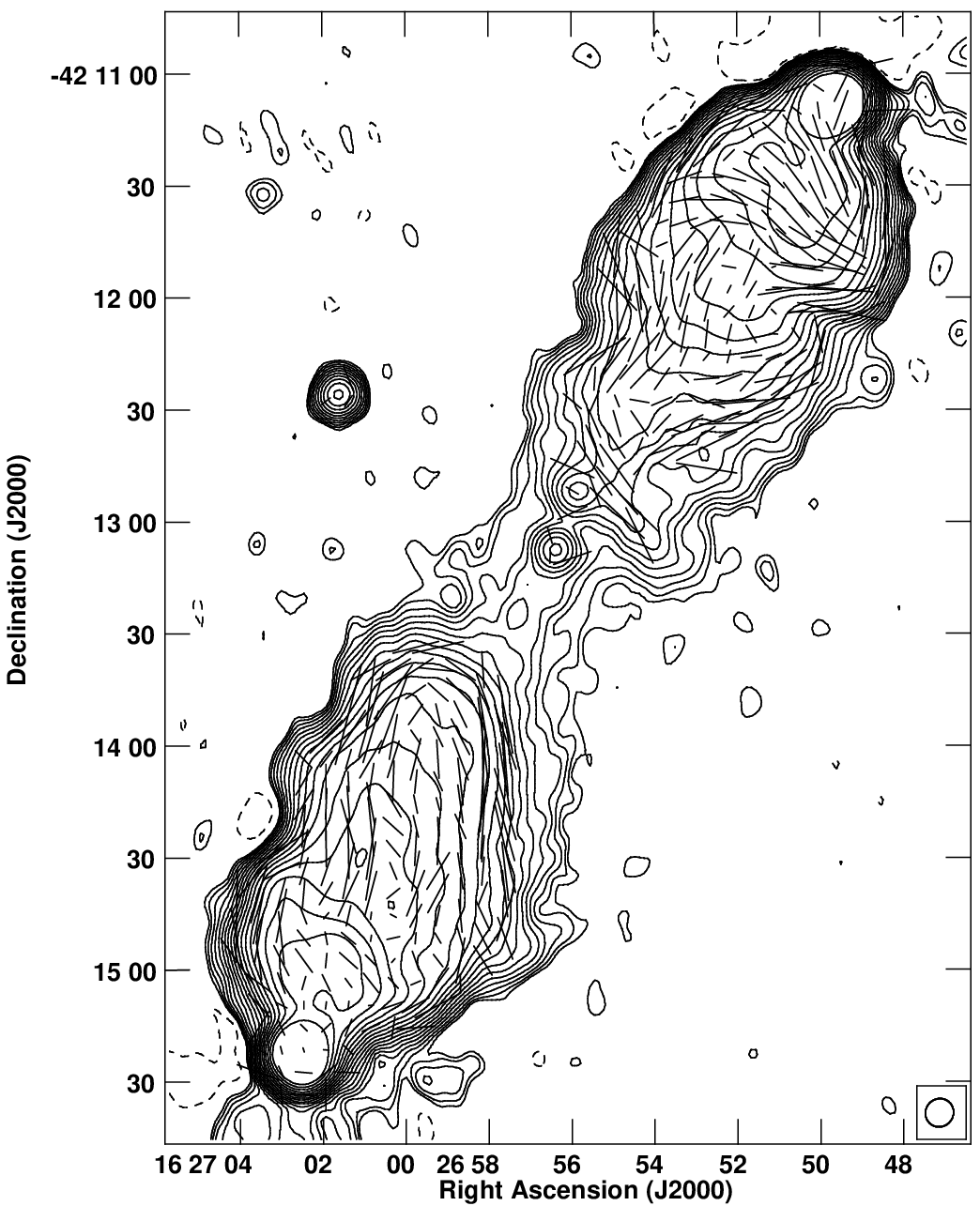}
\caption{J1642563-421307: Contours are at powers of $\sqrt{2}$ from 60
  $\mu$Jy/bm \chga{(lowest, outermost contour)}.  
The vector lengths are proportional to the fractional polarization
shown as ``B'' vectors.
The restoring beam FWHM is given in the box in the lower right corner.}
\label{fig:J164256-421307}
\end{figure}

\noindent
{\bf J1642563-421307:} 
The core of this FRII AGN is at RA(2000)=16 26 56.38, Dec=-42 13 07.6
and the integrated flux density of the source is 1.4 Jy.
A contour plot with superposed polarization ``B'' vectors is given in
Figure \ref{fig:J164256-421307}.
\par
\begin{figure}
\includegraphics[width=3.5in]{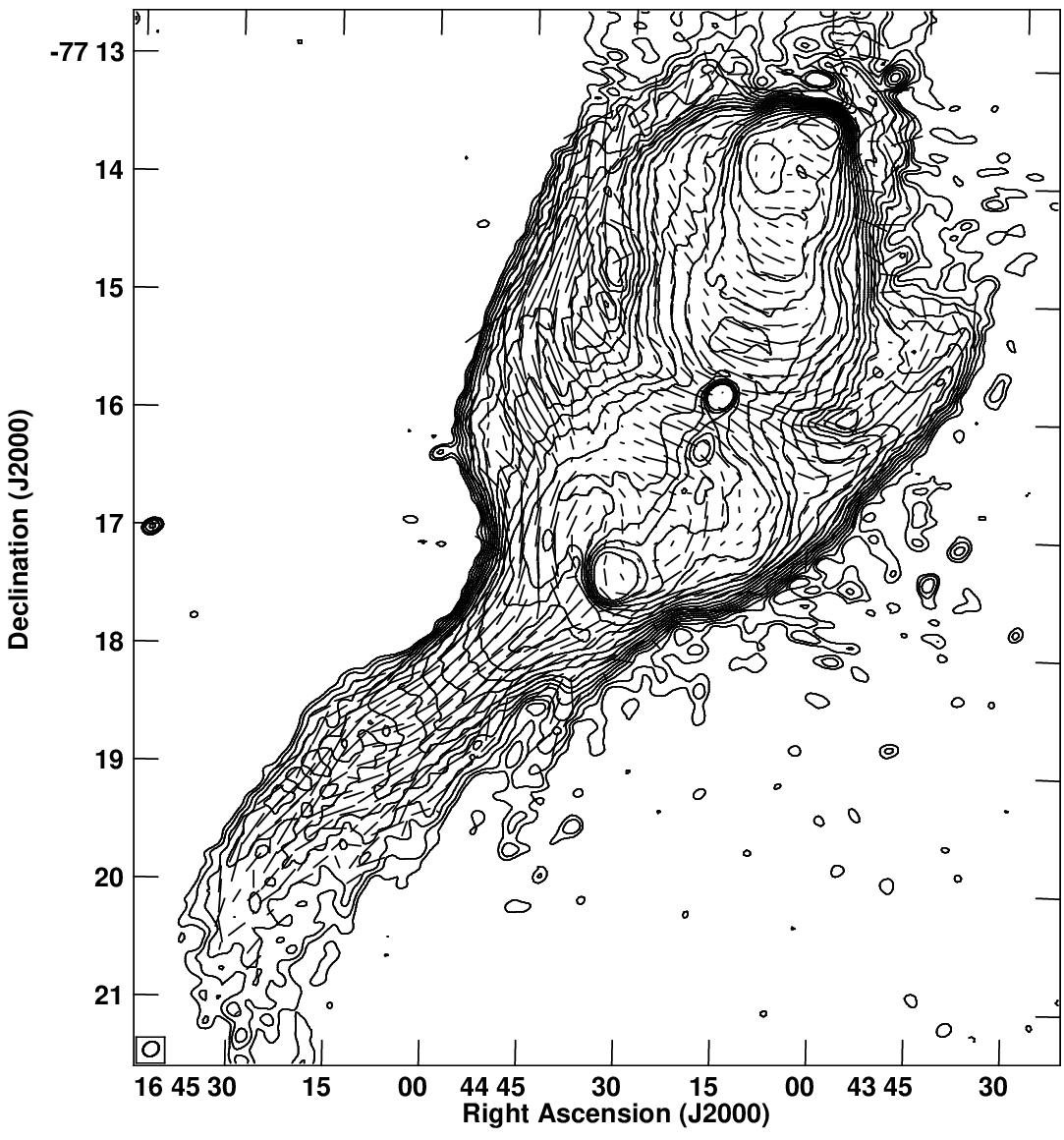}
\caption{J1644158-771549: Contours are at powers of $\sqrt{2}$ from 120
  $\mu$Jy/bm \chga{(lowest, outermost contour)}.  
The vector lengths are proportional to the fractional polarization
shown as ``B'' vectors.
The restoring beam FWHM is given in the box in the lower left corner.}
\label{fig:J164415-771549}
\end{figure}

\noindent
{\bf J1644158-771549:}
The core of this FRII AGN is at RA(2000)=16 44 15.81, Dec=-77 15 49.1
and the source has an integrated flux density of 7.63 Jy.
A contour plot with superposed polarization ``B'' vectors is given in
Figure \ref{fig:J164415-771549}.
The source consists of an inner jet and lobes with extended plumes in
multiple directions; the magnetic field follows these plumes.
The host of this source is WISEA~J164416.12-771549.0 (NED).
\par
\begin{figure}
\includegraphics[width=3.5in]{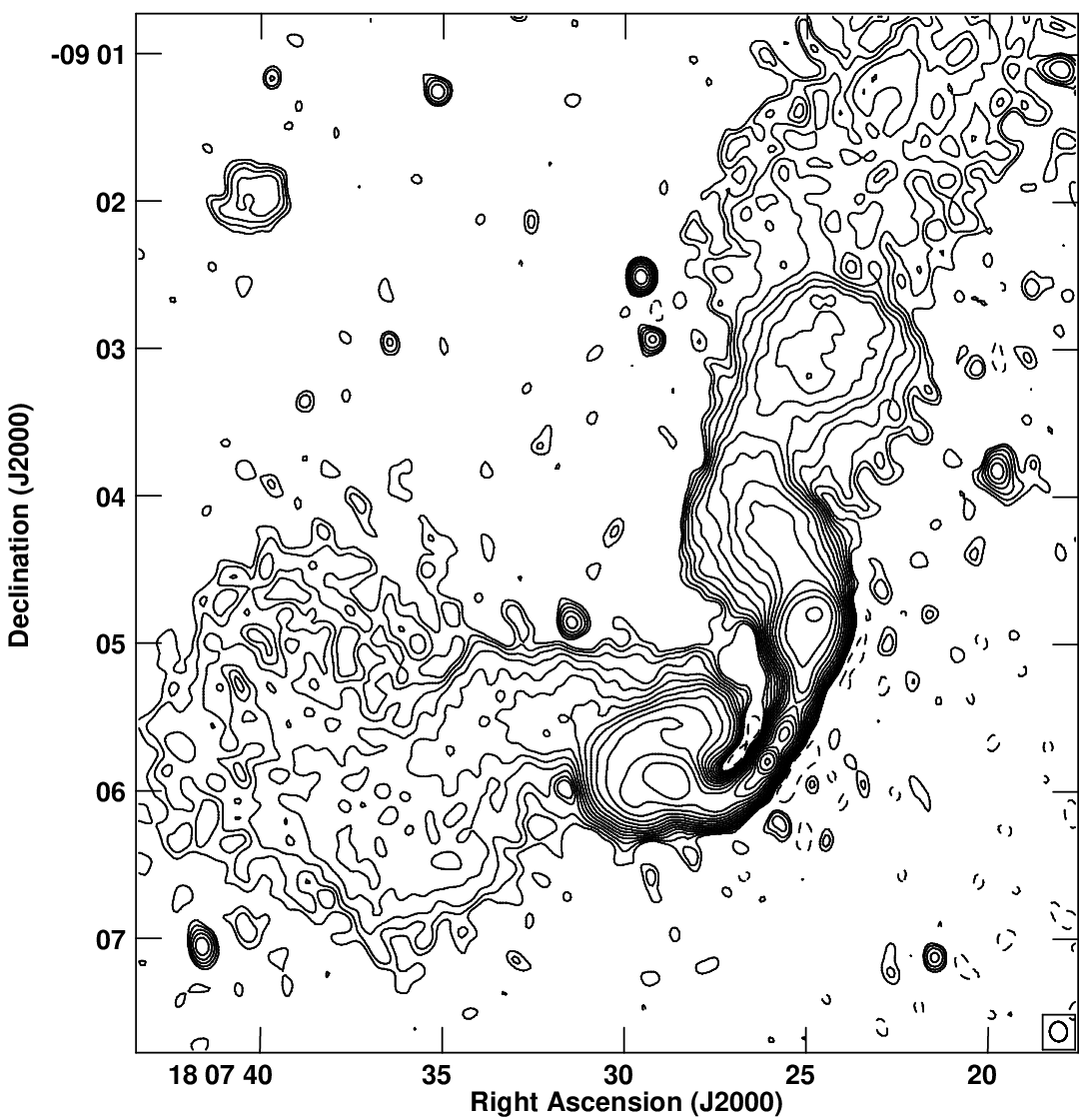}
\caption{J1807256-090536: Contours are at powers of $\sqrt{2}$ from
  74\,$\mu$Jy/bm \chga{(lowest, outermost contour)}.
The restoring beam FWHM is given in the box in the lower right corner.}
\label{fig:J180725-090536}
\end{figure}

\noindent
{\bf J1807256-090536:}
The core of this bent FRI AGN is at RA(2000)=18 07 25.61 Dec = -09 05
36.6 and the integrated flux density of the source is 0.45 Jy.
After several bends the jets expand and drop below the surface
brightness limit of the image.
The source is almost completely depolarized.
A contour plot is shown in Figure \ref{fig:J180725-090536}.
The host galaxy of this source is identitied with
WISEA~J180726.05-090547.1, 2MASX~J18072606-0905469,
2MASS~J18072605-0905473, DENIS~J180726.0-090550 and CGMW~3-1582 by NED.
\par
\begin{figure}
\includegraphics[width=3.5in]{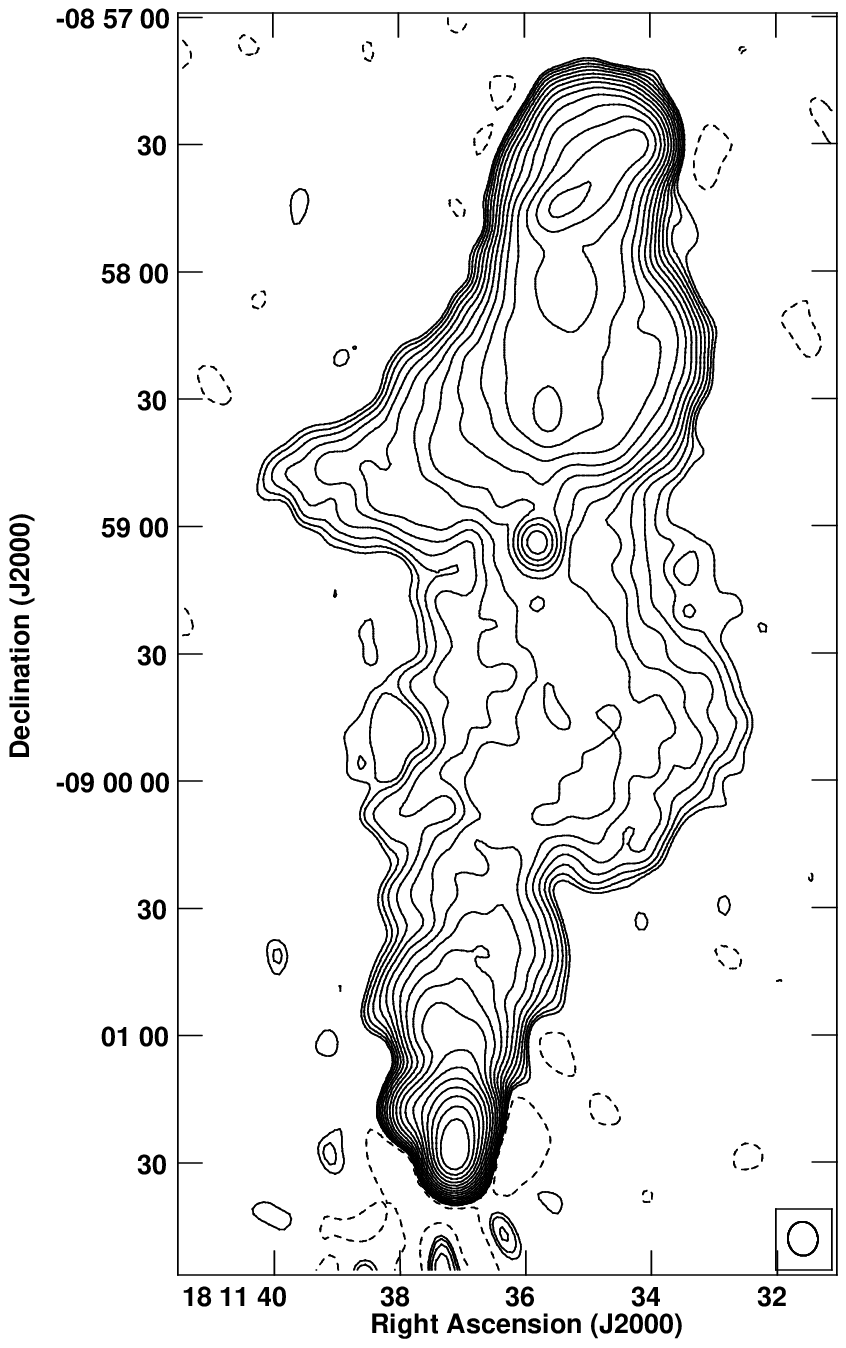}
\caption{J1811358-085903: Contours are at powers of $\sqrt{2}$ from
  74\,$\mu$Jy/bm \chga{(lowest, outermost contour)}.
The restoring beam FWHM is given in the box in the lower right corner.}
\label{fig:J181135-085903}
\end{figure}

\noindent
{\bf J1811358-085903:}
The core of this FRII AGN is at RA(2000)=18 11 35.80, Dec = -08 59
03.9 and the integrated flux density of the source is 0.34 Jy. 
The source is almost completely depolarized;
a contour plot is shown in Figure \ref{fig:J181135-085903}.
The host of this source is 2MASS~J18113592-0858582 (NED).

\subsection{Highly Polarized Sources\label{Hi_Pol}}
Many compact Galactic sources are significantly polarized
which helps distinguish them from the more numerous background
extragalactic sources.
Extended and/or moderately polarized sources are likely to be
background AGN.
Sources with significant Stokes V are very likely Galactic as are
strongly linearly polarized unresolved sources.
The most significantly polarized sources, 50\% or more fractional linear
polarization and 2\% or more circular polarization, from the mosaics are listed in Table
\ref{table:MosaicHiPol}. 
The corresponding highly polarized sources from the single pointing
fields are given in Table \ref{table:OffHiPol}.
The bulk of these sources are likely Galactic, especially if unresolved.

The columns in these tables are 1) the source name, 2) the integrated Stokes I intensity, with error, 3) in--band spectral index, with error, 4) linearly polarized intensity (PPol), with error, 5) Rotation measure (RM), with error, 6) circularly polarized intensity (VPol), with error, 7) ``Y'' or ``N'' indication if the Stokes I image appears to be resolved (a major axis size is given in Table \ref{table:BasicMosaic} or \ref{table:BasicOff} $\Rightarrow$ ``Y'' rather than an upper limit $\Rightarrow$ ``N''), 8) the fractional linear polarization and 9) the fractional circular polarization.
\begin{longrotatetable}
\begin{deluxetable*}{|c|rr|rr|rr|rr|rr|c|r|r|}
  \tablecaption{Highly Polarized Mosaic Compact Sources}
  \tablehead{
    \colhead{Name} & \colhead{IPol} & \colhead{$\pm$} & \colhead{SI} & \colhead{$\pm$} &\colhead{PPol} & \colhead{$\pm$} &
    \colhead{RM} & \colhead{$\pm$} & \colhead{VPol} & \colhead{$\pm$} & 
    \colhead{res\tablenotemark{a}} & \colhead{Fract P\tablenotemark{b}} & \colhead{Fract V\tablenotemark{c}} \\
\colhead{ } & \colhead{$\mu$Jy/bm} & \colhead{} & \colhead{} &\colhead{} &  \colhead{$\mu$Jy/bm} & \colhead{} & \colhead{rad m$^{-2}$} & 
 \colhead{} & \colhead{$\mu$Jy/bm} & \colhead{} & \colhead{} & \colhead{\%} &  \colhead{\%}
}
\startdata
 J1642109-202639 &    129.4 &   19.2  &   \phantom{1.000} & \phantom{1.000}  &   97.0 &  12.8&     24.8 &   1.8&\phantom{1.0}&\phantom{1.0}&  N  &       76.7&\phantom{1.0} \\
 J1645592-182220 &    405.1 &   20.4  &   0.260 & 0.197 &\phantom{1.0}&\phantom{1.0}&   -138.0 &   4.0&   49.4 &   7.2&  N  &\phantom{1.0}&       12.8 \\
 J1651205-215730 &    201.6 &   19.9  &   \phantom{1.000} & \phantom{1.000}  &\phantom{1.0}&\phantom{1.0}&    -18.4 &   3.7&  -37.3 &   9.2&  N  &\phantom{1.0}&      -20.1 \\
 J1653351-224031 &    164.3 &   20.6  &   \phantom{1.000} & \phantom{1.000}  &  110.9 &  14.3&     -4.0 &   8.0&\phantom{1.0}&\phantom{1.0}&  N  &       78.5&\phantom{1.0} \\
 J1653366-212216 &    150.9 &   11.9  &   \phantom{1.000} & \phantom{1.000}  &   81.1 &  14.4&    -10.3 &   7.1&\phantom{1.0}&\phantom{1.0}&  N  &       58.6&\phantom{1.0} \\
 J1654402-224507 &    161.2 &   21.5  &   \phantom{1.000} & \phantom{1.000}  &   87.8 &  16.7&      5.4 &   7.9&\phantom{1.0}&\phantom{1.0}&  N  &       69.2&\phantom{1.0} \\
 J1656480-230901 &    814.0 &   29.3  &  -0.045 & 0.128 &\phantom{1.0}&\phantom{1.0}&     46.8 &   5.1&  -58.7 &  11.5&  N  &\phantom{1.0}&       -7.4 \\
 J1707003-224235 &    158.6 &   14.1  &   \phantom{1.000} & \phantom{1.000}  &   80.7 &  15.2&    -64.3 &   3.6&\phantom{1.0}&\phantom{1.0}&  N  &       52.3&\phantom{1.0} \\
 J1712087-230949 &    243.8 &   13.0  &   0.038 & 0.379 &\phantom{1.0}&\phantom{1.0}&    -25.5 &   5.9&   49.5 &   6.4&  N  &\phantom{1.0}&       20.6 \\
 J1713115-244244 &    372.6 &   52.1  &  -2.030 & 0.348 &  120.5 &  19.4&    -82.1 &   1.2&\phantom{1.0}&\phantom{1.0}&  Y  &       52.4&\phantom{1.0} \\
 J1716297-265516 &    165.0 &   14.8  &   \phantom{1.000} & \phantom{1.000}  &\phantom{1.0}&\phantom{1.0}&\phantom{1.0} & \phantom{1.0} &  109.7 &   8.4&  N  &\phantom{1.0}&       70.1 \\
 J1718203-252734 &    234.7 &   18.8  &  -0.639 & 0.397 &\phantom{1.0}&\phantom{1.0}&\phantom{1.0} & \phantom{1.0} &  -30.5 &   7.2&  N  &\phantom{1.0}&      -14.6 \\
 J1719566-245703 &    284.9 &   17.4  &   1.721 & 0.320 &\phantom{1.0}&\phantom{1.0}&     76.7 &   4.7&  -82.9 &   7.6&  N  &\phantom{1.0}&      -29.1 \\
 J1720125-233357 &    197.5 &   41.1  &   \phantom{1.000} & \phantom{1.000}  &   87.9 &  16.1&    -47.0 &   5.6&\phantom{1.0}&\phantom{1.0}&  N  &       50.1&\phantom{1.0} \\
 J1720184-265207 &    251.9 &   15.5  &  -1.890 & 0.344 &\phantom{1.0}&\phantom{1.0}&    -21.8 &   3.5&  -45.4 &   6.7&  N  &\phantom{1.0}&      -19.8 \\
 J1720341-293315 &     2795 &     85  &  -1.050 & 0.032 &\phantom{1.0}&\phantom{1.0}&      8.5 &   0.9& -160.2 &   8.6&  N  &\phantom{1.0}&       -5.7 \\
 J1720550-272040 &    287.4 &   17.1  &  -0.009 & 0.348 &\phantom{1.0}&\phantom{1.0}&     62.1 &   5.2&  -85.7 &   7.0&  N  &\phantom{1.0}&      -31.9 \\
 J1721016-262647 &     1043 &     34  &  -0.256 & 0.070 &\phantom{1.0}&\phantom{1.0}&\phantom{1.0} & \phantom{1.0} &  -30.5 &   6.9&  N  &\phantom{1.0}&       -2.9 \\
 J1721054-245706 &     1376 &     43  &  -1.129 & 0.049 &\phantom{1.0}&\phantom{1.0}&    -38.1 &   3.3& -162.9 &   7.2&  N  &\phantom{1.0}&      -11.8 \\
 J1723297-282035 &    638.4 &   22.8  &  -1.437 & 0.139 &\phantom{1.0}&\phantom{1.0}&     18.3 &   3.8&  -69.1 &  11.2&  N  &\phantom{1.0}&      -11.0 \\
 J1724163-273043 &    504.6 &   19.3  &   0.799 & 0.161 &\phantom{1.0}&\phantom{1.0}&   -106.1 &   4.2&   83.1 &   7.2&  N  &\phantom{1.0}&       16.5 \\
 J1727284-252820 &    542.1 &   20.1  &   0.998 & 0.156 &\phantom{1.0}&\phantom{1.0}&\phantom{1.0} & \phantom{1.0} &   78.5 &   7.6&  N  &\phantom{1.0}&       14.5 \\
 J1727310-273851 &     1950 &     60  &  -1.460 & 0.049 &\phantom{1.0}&\phantom{1.0}&     -4.3 &   0.6&  118.0 &   7.4&  N  &\phantom{1.0}&        6.2 \\
 J1727479-270710 &    162.7 &   22.2  &   \phantom{1.000} & \phantom{1.000}  &   88.4 &  17.2&   -106.5 &   2.2&\phantom{1.0}&\phantom{1.0}&  N  &       62.4&\phantom{1.0} \\
 J1727520-272724 &    186.2 &   20.7  &   \phantom{1.000} & \phantom{1.000}  &\phantom{1.0}&\phantom{1.0}&     14.9 &   5.0&   39.8 &   7.4&  N  &\phantom{1.0}&       23.2 \\
 J1728202-260816 &    246.5 &   14.9  &  -2.976 & 0.290 &\phantom{1.0}&\phantom{1.0}&    -97.7 &   5.1&   32.4 &   6.7&  N  &\phantom{1.0}&       13.6 \\
 J1730216-230430 &     4419 &    135  &  -0.379 & 0.019 &\phantom{1.0}&\phantom{1.0}&     -1.4 &   0.3&  870.5 &   9.7&  N  &\phantom{1.0}&       19.9 \\
 J1731429-262821 &    946.3 &   45.4  &  -1.596 & 0.122 &\phantom{1.0}&\phantom{1.0}&     -2.3 &   4.6&   41.9 &   9.2&  Y  &\phantom{1.0}&        5.5 \\
 J1732276-302547 &     1360 &     58  &  -0.733 & 0.123 &\phantom{1.0}&\phantom{1.0}&     62.0 &   3.9&   39.0 &   9.5&  Y  &\phantom{1.0}&        5.2 \\
 J1734230-301116 &     2014 &     81  &  -1.046 & 0.086 &\phantom{1.0}&\phantom{1.0}&   -218.3 &   8.0&   40.0 &  10.0&  Y  &\phantom{1.0}&        3.8 \\
 J1739312-234228 &    154.1 &   14.6  &   \phantom{1.000} & \phantom{1.000}  &\phantom{1.0}&\phantom{1.0}&     43.8 &   3.2&  -51.4 &  10.8&  N  &\phantom{1.0}&      -33.7 \\
 J1740067-280533 &    946.6 &   37.6  &  -1.304 & 0.101 &\phantom{1.0}&\phantom{1.0}&     18.2 &   3.5&  237.5 &  10.8&  N  &\phantom{1.0}&       25.4 \\
 J1741013-273355 &     2386 &     89  &  -2.218 & 0.040 &\phantom{1.0}&\phantom{1.0}&    -88.4 &   1.1&  -79.6 &   9.9&  Y  &\phantom{1.0}&       -3.5 \\
 J1742465-272940 &    296.6 &   27.3  &  -0.971 & 0.303 &\phantom{1.0}&\phantom{1.0}&    -21.3 &   3.3&  -49.9 &  11.7&  N  &\phantom{1.0}&      -16.7 \\
 J1743155-315305 &    906.4 &   42.0  &  -0.337 & 0.122 &\phantom{1.0}&\phantom{1.0}&     74.1 &   1.2& -139.3 &  10.8&  Y  &\phantom{1.0}&      -17.3 \\
 J1743367-315021 &     2458 &     80  &  -2.586 & 0.036 &\phantom{1.0}&\phantom{1.0}&   -218.0 &   0.8&   64.2 &  10.2&  N  &\phantom{1.0}&        2.6 \\
 J1744086-242452 &     1487 &     57  &  -2.209 & 0.100 &\phantom{1.0}&\phantom{1.0}&   -106.0 &   1.4&  198.6 &  10.7&  Y  &\phantom{1.0}&       15.9 \\
 J1744331-260813 &    203.9 &   17.4  &   \phantom{1.000} & \phantom{1.000}  &\phantom{1.0}&\phantom{1.0}&    -53.7 &   5.6&   40.0 &   9.1&  N  &\phantom{1.0}&       19.5 \\
 J1747180-330916 &     1056 &     37  &  -2.098 & 0.085 &\phantom{1.0}&\phantom{1.0}&    147.6 &   3.5& -200.9 &   8.4&  N  &\phantom{1.0}&      -19.2 \\
 J1748045-244634 &     1172 &     53  &   \phantom{1.000} & \phantom{1.000}  &\phantom{1.0}&\phantom{1.0}&    108.9 &   2.4&  -79.2 &  15.9&  Y  &\phantom{1.0}&       -7.5 \\
 J1749135-300235 &     4193 &    154  &  -2.077 & 0.038 &\phantom{1.0}&\phantom{1.0}&   -137.0 &   0.6& -454.7 &  15.8&  Y  &\phantom{1.0}&      -11.5 \\
 J1750101-245826 &    203.5 &   28.2  &   \phantom{1.000} & \phantom{1.000}  &  110.3 &  20.5&     30.1 &   4.1&\phantom{1.0}&\phantom{1.0}&  N  &       55.9&\phantom{1.0} \\
 J1750473-315743 &     1760 &     54  &  -1.482 & 0.052 &\phantom{1.0}&\phantom{1.0}&     74.5 &   0.5&   57.7 &   7.2&  N  &\phantom{1.0}&        3.3 \\
 J1751068-321827 &    581.1 &   21.0  &   0.777 & 0.196 &\phantom{1.0}&\phantom{1.0}&     31.7 &   4.4&  101.0 &   6.8&  N  &\phantom{1.0}&       17.8 \\
 J1751159-294351 &    915.5 &   53.9  &  -1.417 & 0.101 &\phantom{1.0}&\phantom{1.0}&    -33.5 &   2.3&   58.6 &  11.5&  N  &\phantom{1.0}&        6.5 \\
 J1751296-284456 &    150.1 &   31.6  &   \phantom{1.000} & \phantom{1.000}  &   91.6 &  16.8&     35.0 &   4.2&\phantom{1.0}&\phantom{1.0}&  N  &       72.7&\phantom{1.0} \\
 J1751326-285746 &    532.3 &   31.0  &   \phantom{1.000} & \phantom{1.000}  &  312.0 &  16.8&     36.5 &   1.3&\phantom{1.0}&\phantom{1.0}&  N  &       58.5&\phantom{1.0} \\
 J1752290-284904 &     1177 &     45  &  -0.713 & 0.083 &\phantom{1.0}&\phantom{1.0}&    -63.1 &   2.8&   80.9 &  12.8&  N  &\phantom{1.0}&        7.1 \\
 J1752586-280637 &    46720 &   1402  &   \phantom{1.000} & \phantom{1.000}  &\phantom{1.0}&\phantom{1.0}&     96.0 &   0.1& -1223.0 &  14.5&  N  &\phantom{1.0}&       -2.6 \\
 J1753064-293000 &    370.7 &   32.4  &  -2.364 & 0.254 &\phantom{1.0}&\phantom{1.0}&    -79.3 &   3.9&   59.8 &   9.7&  N  &\phantom{1.0}&       17.1 \\
 J1753065-285125 &    254.2 &   26.6  &  -2.072 & 0.326 &\phantom{1.0}&\phantom{1.0}&     -0.7 &   3.8&   87.7 &  10.9&  N  &\phantom{1.0}&       34.1 \\
 J1753390-333127 &    160.4 &   13.4  &   \phantom{1.000} & \phantom{1.000}  &\phantom{1.0}&\phantom{1.0}&    -53.3 &   3.5&   29.3 &   6.6&  N  &\phantom{1.0}&       18.4 \\
 J1755165-313625 &    188.3 &   14.8  &   \phantom{1.000} & \phantom{1.000}  &\phantom{1.0}&\phantom{1.0}&\phantom{1.0} & \phantom{1.0} &  -28.7 &   7.1&  N  &\phantom{1.0}&      -16.5 \\
 J1755248-335830 &    170.5 &   12.6  &   \phantom{1.000} & \phantom{1.000}  &\phantom{1.0}&\phantom{1.0}&    222.7 &   2.9&   57.3 &   6.6&  N  &\phantom{1.0}&       36.0 \\
 J1756214-312203 &    10381 &    357  &  -0.947 & 0.011 &\phantom{1.0}&\phantom{1.0}&    265.6 &   1.2& -219.1 &  26.6&  Y  &\phantom{1.0}&       -2.6 \\
 J1759205-301547 &    154.9 &   11.3  &   \phantom{1.000} & \phantom{1.000}  &\phantom{1.0}&\phantom{1.0}&   -132.1 &   4.5&   33.4 &   7.1&  N  &\phantom{1.0}&       21.5 \\
 J1759220-310721 &     1404 &     44  &  -1.458 & 0.066 &\phantom{1.0}&\phantom{1.0}&    142.4 &   1.1&  -47.7 &   7.1&  N  &\phantom{1.0}&       -3.5 \\
 J1759482-292207 &    887.7 &   29.5  &  -2.580 & 0.101 &\phantom{1.0}&\phantom{1.0}&    200.0 &   1.6&   42.2 &   7.9&  N  &\phantom{1.0}&        4.8 \\
 J1759532-262424 &    162.3 &   22.9  &  -2.067 & 1.175 &   93.8 &  16.0&     20.7 &   3.1&\phantom{1.0}&\phantom{1.0}&  N  &       56.8&\phantom{1.0} \\
 J1800123-274431 &    168.6 &   17.5  &   \phantom{1.000} & \phantom{1.000}  &\phantom{1.0}&\phantom{1.0}&\phantom{1.0} & \phantom{1.0} &   82.4 &   9.4&  N  &\phantom{1.0}&       53.4 \\
 J1800321-273535 &    386.1 &   18.9  &   \phantom{1.000} & \phantom{1.000}  &\phantom{1.0}&\phantom{1.0}&     53.3 &   5.5&  -51.1 &   9.2&  N  &\phantom{1.0}&      -13.8 \\
 J1801468-292038 &     3190 &     96  &  -1.789 & 0.031 &\phantom{1.0}&\phantom{1.0}&    -57.3 &   0.2&  227.5 &   7.2&  N  &\phantom{1.0}&        7.1 \\
 J1803316-271204 &     1581 &     53  &  -1.367 & 0.065 &\phantom{1.0}&\phantom{1.0}&   -164.8 &   0.6&   92.6 &  11.7&  N  &\phantom{1.0}&        6.0 \\
 J1803351-300159 &    841.1 &   33.2  &  -2.131 & 0.120 &\phantom{1.0}&\phantom{1.0}&\phantom{1.0} & \phantom{1.0} &  -37.6 &   7.2&  Y  &\phantom{1.0}&       -5.1 \\
 J1803444-332910 &    949.0 &   29.9  &  -1.206 & 0.059 &\phantom{1.0}&\phantom{1.0}&    -33.8 &   1.9&  -77.5 &   5.9&  N  &\phantom{1.0}&       -8.2 \\
 J1804015-285846 &     1263 &     39  &  -2.402 & 0.078 &\phantom{1.0}&\phantom{1.0}&     94.3 &   2.4&  105.0 &   6.8&  N  &\phantom{1.0}&        8.4 \\
 J1804211-271731 &    859.6 &   35.3  &   \phantom{1.000} & \phantom{1.000}  &\phantom{1.0}&\phantom{1.0}&     40.7 &   1.0& -105.2 &  11.6&  N  &\phantom{1.0}&      -12.5 \\
 J1805264-292953 &    197.7 &   13.6  &   \phantom{1.000} & \phantom{1.000}  &\phantom{1.0}&\phantom{1.0}&    -95.2 &   3.5&  132.6 &   6.7&  N  &\phantom{1.0}&       67.1 \\
 J1806589-345316 &    132.3 &   15.9  &   \phantom{1.000} & \phantom{1.000}  &   67.9 &  13.2&     27.8 &   6.1&\phantom{1.0}&\phantom{1.0}&  N  &       53.4&\phantom{1.0} \\
 J1807084-271502 &    935.4 &   34.2  &   \phantom{1.000} & \phantom{1.000}  &\phantom{1.0}&\phantom{1.0}&     37.6 &   9.4&  -61.9 &  14.5&  N  &\phantom{1.0}&       -6.6 \\
 J1808045-324932 &     1666 &     51  &  -1.102 & 0.034 &\phantom{1.0}&\phantom{1.0}&    295.9 &   1.7&  111.6 &   6.3&  N  &\phantom{1.0}&        6.7 \\
 J1811372-320650 &    347.1 &   14.0  &  -0.388 & 0.169 &\phantom{1.0}&\phantom{1.0}&     90.7 &   3.9&   60.0 &   6.7&  N  &\phantom{1.0}&       17.4 \\
 J1813023-310112 &    246.3 &   12.3  &  -1.128 & 0.232 &\phantom{1.0}&\phantom{1.0}&     58.5 &   5.4&   28.0 &   7.0&  N  &\phantom{1.0}&       11.6 \\
 J1813324-281403 &     1752 &     54  &  -1.116 & 0.032 &\phantom{1.0}&\phantom{1.0}&    -58.3 &   1.6&  127.6 &   9.4&  N  &\phantom{1.0}&        7.3 \\
 J1813457-331737 &    145.4 &   15.6  &   \phantom{1.000} & \phantom{1.000}  &\phantom{1.0}&\phantom{1.0}&    102.6 &   4.4&  144.4 &   6.8&  N  &\phantom{1.0}&      100.8 \\
 J1816343-323411 &    447.1 &   16.8  &  -1.068 & 0.132 &\phantom{1.0}&\phantom{1.0}&\phantom{1.0} & \phantom{1.0} &   33.7 &   6.0&  N  &\phantom{1.0}&        7.5 \\
 J1817027-305010 &    142.5 &   20.8  &   \phantom{1.000} & \phantom{1.000}  &   71.6 &  14.1&     79.9 &   3.3&\phantom{1.0}&\phantom{1.0}&  N  &       59.1&\phantom{1.0} \\
 J1817098-331730 &    143.9 &   11.6  &   \phantom{1.000} & \phantom{1.000}  &\phantom{1.0}&\phantom{1.0}&\phantom{1.0} & \phantom{1.0} &  -34.3 &   5.9&  N  &\phantom{1.0}&      -25.9 \\
 J1818310-325417 &    210.8 &   13.1  &   \phantom{1.000} & \phantom{1.000}  &\phantom{1.0}&\phantom{1.0}&     99.2 &   8.5&   85.0 &   6.2&  N  &\phantom{1.0}&       42.0 \\
 J1819522-291633 &    262.2 &   16.3  &   0.104 & 0.433 &\phantom{1.0}&\phantom{1.0}&\phantom{1.0} & \phantom{1.0} &  -52.9 &  11.0&  N  &\phantom{1.0}&      -20.5 \\
 J1827317-325237 &     1329 &     42  &  -1.391 & 0.044 &\phantom{1.0}&\phantom{1.0}&     39.4 &   1.9&  -47.0 &   8.6&  N  &\phantom{1.0}&       -3.6 \\
 J1829520-335540 &    196.1 &   11.3  &   \phantom{1.000} & \phantom{1.000}  &\phantom{1.0}&\phantom{1.0}&     21.1 &   4.8&   75.4 &   6.3&  N  &\phantom{1.0}&       38.9 \\
 J1829529-334304 &    165.4 &   10.3  &   \phantom{1.000} & \phantom{1.000}  &\phantom{1.0}&\phantom{1.0}&     17.1 &   5.6&   73.7 &   6.3&  N  &\phantom{1.0}&       45.2 \\
 J1841547-360748 &    200.4 &   27.1  &   \phantom{1.000} & \phantom{1.000}  &\phantom{1.0}&\phantom{1.0}&    -13.0 &   3.5&  -50.7 &   6.5&  Y  &\phantom{1.0}&      -38.2 \\
 J1843583-355909 &    287.6 &   13.2  &   1.270 & 0.209 &\phantom{1.0}&\phantom{1.0}&\phantom{1.0} & \phantom{1.0} &   30.6 &   5.8&  N  &\phantom{1.0}&       10.9 \\
 J1852260-373037 &    178.8 &   14.7  &   \phantom{1.000} & \phantom{1.000}  &\phantom{1.0}&\phantom{1.0}&      8.9 &   5.2&   73.9 &   8.6&  N  &\phantom{1.0}&       41.5 \\
 J1853059-361022 &    446.8 &   18.2  &   0.232 & 0.129 &\phantom{1.0}&\phantom{1.0}&     -9.9 &   3.4&  -34.5 &   7.4&  N  &\phantom{1.0}&       -7.9 \\
 J1857451-371937 &    505.9 &   18.2  &   0.166 & 0.106 &\phantom{1.0}&\phantom{1.0}&     19.8 &   9.4& -140.6 &   6.2&  N  &\phantom{1.0}&      -27.8 \\
 J1901480-365721 &     1048 &     39  &   0.588 & 0.079 &\phantom{1.0}&\phantom{1.0}&     13.7 &   6.5&  100.6 &   5.8&  Y  &\phantom{1.0}&       14.1 \\
 J1903063-371641 &    149.9 &    9.3  &   \phantom{1.000} & \phantom{1.000}  &\phantom{1.0}&\phantom{1.0}&    -50.1 &   4.5&   65.6 &   5.7&  N  &\phantom{1.0}&       43.7 \\
 J1906256-370348 &    258.2 &   20.8  &   \phantom{1.000} & \phantom{1.000}  &\phantom{1.0}&\phantom{1.0}&     12.5 &   3.5&  153.4 &   6.4&  Y  &\phantom{1.0}&       78.1 \\
 J1700277-220742 &     74.1 &   11.0  &   \phantom{1.000} & \phantom{1.000}  &\phantom{1.0}&\phantom{1.0}&\phantom{1.0} & \phantom{1.0} &  -52.5 &   6.5&  N  &\phantom{1.0}&      -69.5 \\
 J1711229-243850 &    191.5 &   34.3  &   \phantom{1.000} & \phantom{1.000}  &\phantom{1.0}&\phantom{1.0}&\phantom{1.0} & \phantom{1.0} &  -62.4 &   7.0&  N  &\phantom{1.0}&      -33.2 \\
 J1901405-364431 &     78.1 &    8.4  &   \phantom{1.000} & \phantom{1.000}  &\phantom{1.0}&\phantom{1.0}&\phantom{1.0} & \phantom{1.0} &   50.7 &   5.8&  N  &\phantom{1.0}&       64.4 \\
\enddata
\tablenotetext{a}{Is the source resolved? ``Y'' or ``N''}
\tablenotetext{b}{Fractional linear polarization, min (P/I)=50\%.}
\tablenotetext{c}{Fractional circular polarization, min($|$V$|$)/I=2\%.}
\label{table:MosaicHiPol}
\end{deluxetable*}
\end{longrotatetable}

\begin{longrotatetable}
\begin{deluxetable*}{|c|rr|rr|rr|rr|rr|c|r|r|}
  \tablecaption{Highly Polarized Single Pointing Compact Sources}
  \tablehead{
    \colhead{Name} & \colhead{IPol} & \colhead{$\pm$} & \colhead{SI} & \colhead{$\pm$} &\colhead{PPol} & \colhead{$\pm$} &
    \colhead{RM} & \colhead{$\pm$} & \colhead{VPol} & \colhead{$\pm$} & 
    \colhead{res\tablenotemark{a}} & \colhead{Fract P\tablenotemark{b}} & \colhead{Fract V\tablenotemark{c}} \\
\colhead{ } & \colhead{$\mu$Jy/bm} & \colhead{} & \colhead{} &\colhead{} &  \colhead{$\mu$Jy/bm} & \colhead{} & \colhead{rad m$^{-2}$} & 
 \colhead{} & \colhead{$\mu$Jy/bm} & \colhead{} & \colhead{} & \colhead{\%} &  \colhead{\%}
}
\startdata
J1624381-421413 &    121.6 &   26.2  &   \phantom{1.000} & \phantom{1.000}  &   110.7 &  21.7&     13.3 &   6.2&\phantom{1.0}&\phantom{1.0}&  N  &      101.3&\phantom{1.0} \\
 J1628472-415239 &     3161 &     97  &   1.297 & 0.024 & \phantom{1.0}&\phantom{1.0}&      1.7 &   2.4& -467.7 &  15.4&  N  &\phantom{1.0}&      -15.1 \\
 J1651075-764240 &     1204 &     40  &  -1.327 & 0.050 & \phantom{1.0}&\phantom{1.0}&     25.8 &   0.7&   84.9 &   9.4&  N  &\phantom{1.0}&        7.1 \\
 J1709484-481519 &     1950 &     70  &  -0.733 & 0.042 & \phantom{1.0}&\phantom{1.0}&   -147.6 &   2.1&  -42.2 &  10.4&  Y  &\phantom{1.0}&       -2.8 \\
 J1720572-021223 &     1463 &     49  &  -2.770 & 0.065 & \phantom{1.0}&\phantom{1.0}&      6.2 &   0.9&  121.9 &   9.9&  N  &\phantom{1.0}&        8.4 \\
 J1734518-511530 &     2382 &     84  &  -0.567 & 0.031 & \phantom{1.0}&\phantom{1.0}&     -2.3 &   0.9&   42.9 &   8.6&  Y  &\phantom{1.0}&        2.3 \\
 J1734550-512832 &     3052 &    110  &  -0.656 & 0.022 & \phantom{1.0}&\phantom{1.0}&    -64.5 &   1.0&   56.7 &  10.9&  Y  &\phantom{1.0}&        2.2 \\
 J1837524-553140 &    301.0 &   17.0  &  -0.177 & 0.166 & \phantom{1.0}&\phantom{1.0}&    -48.4 &   5.7&   35.6 &   8.9&  N  &\phantom{1.0}&       12.5 \\
 J1838363-554454 &     2444 &     90  &  -0.265 & 0.021 & \phantom{1.0}&\phantom{1.0}&     51.4 &   3.1&  -59.8 &  14.5&  Y  &\phantom{1.0}&       -2.6 \\
 J1838467+152325 &    561.9 &   56.9  &   \phantom{1.000} & \phantom{1.000}  & \phantom{1.0}&\phantom{1.0}&   -138.3 &   3.4&   79.0 &  11.2&  Y  &\phantom{1.0}&       21.5 \\
 J1845009-140905 &     1430 &     58  &  -1.755 & 0.048 & \phantom{1.0}&\phantom{1.0}&    -12.9 &   4.3&  967.2 &  14.8&  Y  &\phantom{1.0}&       76.4 \\
 J1930323+082925 &    123.5 &   23.4  &   \phantom{1.000} & \phantom{1.000}  &    63.3 &  11.8&    -66.8 &   8.7&\phantom{1.0}&\phantom{1.0}&  N  &       55.6&\phantom{1.0} \\
 J1949139+055306 &    219.0 &   14.8  &  -0.861 & 0.277 & \phantom{1.0}&\phantom{1.0}&      2.4 &   5.4&  159.5 &   7.8&  N  &\phantom{1.0}&       74.6 \\
 J1949170+052733 &    176.5 &   29.6  &   \phantom{1.000} & \phantom{1.000}  &    79.7 &  15.9&    -91.7 &   6.6&\phantom{1.0}&\phantom{1.0}&  N  &       56.8&\phantom{1.0} \\
 J2023032+005824 &    181.3 &   22.3  &   \phantom{1.000} & \phantom{1.000}  & \phantom{1.0}&\phantom{1.0}&     -2.2 &   1.9&   99.0 &   9.7&  N  &\phantom{1.0}&       59.3 \\
\enddata
\tablenotetext{a}{Is the source resolved? ``Y'' or ``N''}
\tablenotetext{b}{Fractional linear polarization, min (P/I)=50\%.}
\tablenotetext{c}{Fractional circular polarization, min($|$V$|$)/I=2\%.}
\label{table:OffHiPol}
\end{deluxetable*}
\end{longrotatetable}

\subsubsection{Individual Polarized Sources}
An extensive examination of the highly polarized sources from the
mosaic fields is given in \cite{Frail2024}.
The following discusses some of the more highly polarized sources from
the single pointing fields shown in Table \ref{table:OffHiPol} which were
matched in either SIMBAD (http://simbad.cds.unistra.fr/simbad/) or NED
(https://ned.ipac.caltech.edu/).
\begin{itemize}
\item {\bf J1628472-415239}
SIMBAD matches this source with VASC~J1628-41, an RS CVn Variable of
spectral type K3III with parallax 3.49 mas.
\item {\bf J1651075-764240}
SIMBAD matches this source with pulsar PSR~J1651-7642.
\item {\bf J1709484-481519}
NED matches this source with WISEA J170949.19-481513.3 and 2MASS J17094920-4815132.
 \item {\bf J1720572-021223}
SIMBAD matches this source with pulsar PSR~B1718-02.
\item {\bf J1734518-511530}
NED matches this source with 2MASS~J17345199-5115283.
\item {\bf J1734550-512832}
NED matches this source with WISEA J173455.81-512827.0.
\item {\bf J1837524-553140}
NED matches this source with WISEA J183752.64-553142.6 and 2MASS J18375264-5531426.
\item {\bf J1838363-554454}
NED matches this source with WISEA~J183836.56-554454.9 
\item {\bf J1838467+152325}
SIMBAD matches this source with pulsar PSR~J1838+1523.
\item {\bf J1845009-140905}
SIMBAD  matches this source with UCAC3~152-281185, a high proper
motion star of spectral type M5 and parallax 55.12 mas.
\item {\bf J1930323+082925}
SIMBAD lists ZTF J193032.23+082913.2 an RS CVn Variable 12\asec\ away.  NED gives a closer match (3\asec) with WISEA J193032.46+082927.4 and 2MASS J19303245+0829279. 
\item {\bf J1949139+055306}
SIMBAD  matches this source with Gaia~DR3~4296795272753738496, a high
proper motion star with parallax 28.5 mas.
\item {\bf J2023032+005824}
SIMBAD matches this source with Gaia DR3 4230875290045369344, a star.
\end{itemize}

\section{Search for Giant Radio Galaxies\label{GRGs}}
The search for Giant Radio Galaxies (GRGs) was conducted using the mosaics described in Section \ref{Mosaics}.
\chg{GRGs are a rare class of radio galaxies with projected linear sizes exceeding 0.7 Mpc, making them among the largest known structures in the universe. While most radio galaxies do not grow to such enormous scales, the factors driving their extreme growth remain uncertain, making GRGs key subjects of interest for studying galaxy evolution, jet propagation, and the role of the intergalactic medium.} This comprehensive analysis involved identifying 1734 extended radio sources and assessing their potential as GRGs based on their angular extents, as described in greater detail in \cite{PreetThesis2023}.

The angular sizes of these sources were measured using the OBIT software \citep{OBIT}, where the structures of the radio lobes were manually traced using piecewise linear sections, following their contours, including bends, to accurately capture their full extent. The locations of potential nuclei in the host galaxies were identified through visual inspection. Radio core positions were matched with probable host galaxies within a 10-arcsecond radius in the 2MASS \citep{2MASS} and allWISE \citep{allWISE} catalogs. To refine host galaxy identification, candidates with separations greater than 2 arcseconds between 2MASS and allWISE data were excluded.

The identified host positions were cross-referenced with the NASA Extragalactic Database (NED \url{https://ned.ipac.caltech.edu/}),
yielding redshift values for 22 sources. For these sources, distances and physical sizes were calculated using the NED redshifts. When redshift data was not available, estimates were derived from the K-band magnitudes of the host galaxies, leveraging known correlations between K-band luminosities and stellar mass. This approach provided lower-limit estimates for distances, which were especially valuable for higher-redshift galaxies where direct measurements were lacking.

Double Radio-source Associated with Active Galactic Nucleus (DRAGNs), which are often hosted by massive elliptical galaxies, served as a key focus for this study. The K-band emission from these galaxies, dominated by an older stellar population, acts as a reliable proxy for total stellar mass (\cite{Merloni_2010}). Reliable 2MASS K-band measurements were utilized when available; otherwise, WISE observations were used to estimate K-band magnitudes from the W1 and W2 bands. The K-band magnitudes were corrected for Galactic extinction \citep{Schlafly_2011} (\url{https://irsa.ipac.caltech.edu/applications/DUST/}). The average absolute K-band magnitude of host galaxies with measured redshifts in our sample was determined to be $-24.3\pm0.9$ \citep{PreetThesis2023}, and this value was used to estimate redshifts ($z_{calc}$), distances ($D_A$), and physical sizes ($I_A$)\footnote{The cosmology assumed was
  H$_0$=67.4, $\Omega_M$=0.3142 and $\Omega_{\lambda}$=0.717}.

\begin{figure}[!ht]
    \centering
    \includegraphics[width=\linewidth]{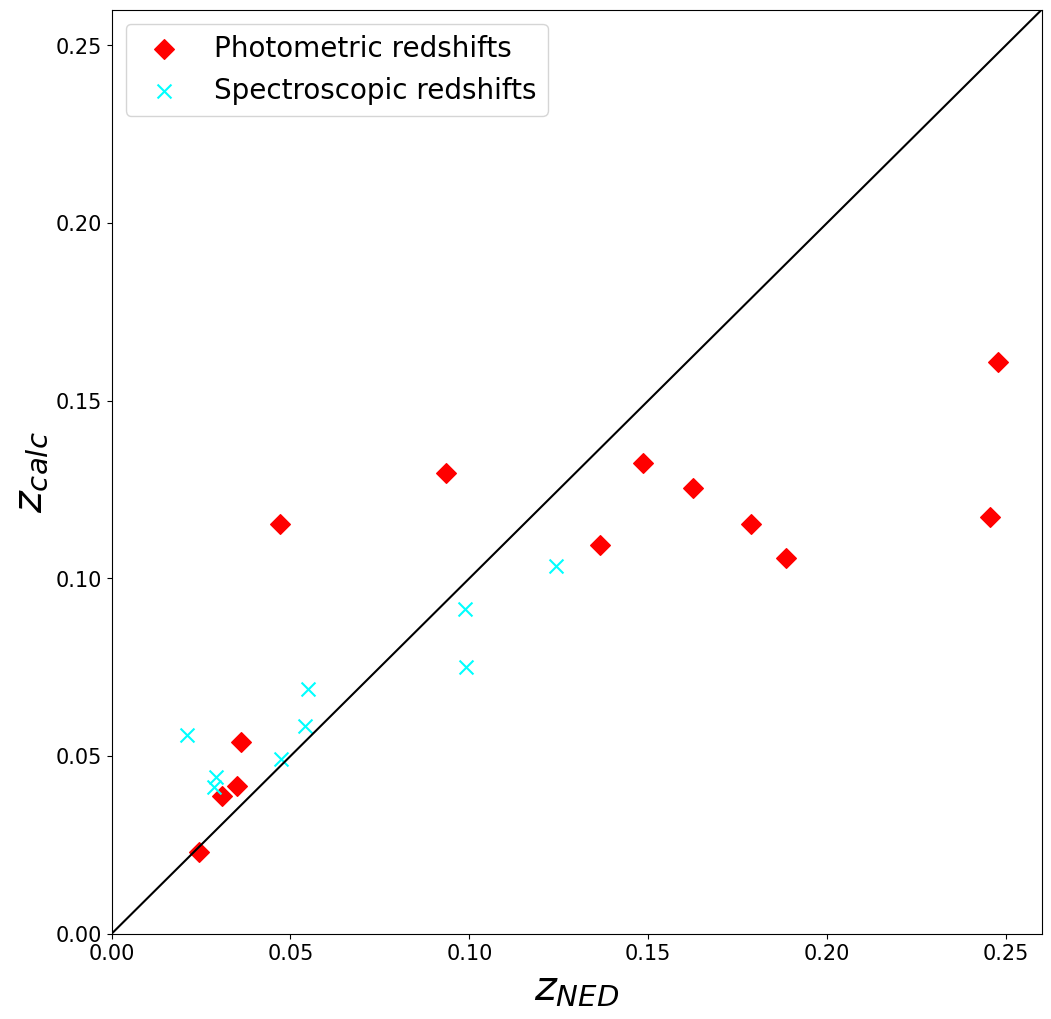}
    \caption{Comparison between $z_{NED}$ and $z_{calc}$ for sources with available NED redshifts (see Table \ref{table:ExtendedCat}), showing a close correlation up to $z \approx 0.1$ and underestimation for higher redshifts. The line gives the 1:1 equality.}
    \label{z_vs_z}
\end{figure}

Table \ref{table:ExtendedCat} presents the extended radio sources with available NED redshifts, providing a comparison between $z_{NED}$ and $z_{calc}$. This table, along with Figure \ref{z_vs_z}, demonstrates that redshift estimates based on K-band magnitudes align closely with NED values up to approximately $z \approx 0.1$, but show underestimation at higher redshifts. The distances derived from these redshifts ($D_z$ for NED-based and $D_A$ for K-band estimates) were used to calculate the physical extents of the sources. Consequently, the distances and projected linear sizes of these sources are considered lower limits. Although the calculated redshifts carry inherent biases and uncertainties, the sources identified as GRGs, very likely, are GRGs.

\chg{Based on the criterion that their projected physical sizes must be 0.7 Mpc or greater, 17 probable GRGs were identified in this study. These sizes were determined using the measured angular extents of the sources along with their redshift estimates, either from NED ($z_{NED}$) or derived from K-band magnitudes ($z_{calc}$). This classification ensured that only the most extended radio galaxies were included in the final selection.}
These GRGs are detailed in Table \ref{table:GRG_list}. 
These methods ensured reliable identification despite the limitations at higher redshifts, where the assumption of a consistent absolute K-band magnitude may not hold due to galaxy evolution.

\chg{To account for uncertainties in the estimated projected physical sizes of the sources, a Monte Carlo analysis was performed. The uncertainty in the assumed absolute K-band magnitude of the host galaxies introduced errors in the distance modulus, which propagated into the redshift, distance, and ultimately, the physical size estimations. To quantify the impact of these uncertainties, we generated 100,000 realizations of the dataset, where each source’s projected physical size was varied by drawing random values from a normal distribution centered on its estimated size with a standard deviation equal to its calculated uncertainty. For each realization, the number of sources exceeding the 0.7 Mpc threshold was counted, yielding a distribution of probable GRG counts. The mean and standard deviation of this distribution resulted in an estimated GRG count of $27.9 \pm 3.9$ over an area of approximately 173 square deg. }

\chg{A comparison of the 17 identified GRGs and the MC estimate of
$27.9 \pm 3.9$  is not straightforward.
Due to the lack of measured redshifts, the distance estimates relied
on the assumption that all host galaxies had the same absolute K band
magnitude.
Figure \ref{z_vs_z} shows this not to be the case and the distances
beyond a z$\approx$0.1 are significantly underestimated and thus the
linear sizes are underestimated.
Objects identified as GRGs very likely are but many will likely be
missed. 
17 GRGs should be taken as the lower limit.
   The MC analysis was based on the erroneous assumption that all the
uncertainties are known, Gaussian and zero mean.
The distances are particularly troublesome and their errors are very
nongaussian and biased.  
The MC value is an estimate of what might have been measured had the
average K band absolute magnitude of the host galaxies been constant
with cosmic epoch.  
This estimate is more of an upper bound.
}

A comprehensive list of the large angular-sized radio sources is available online (via \url{https://doi.org/10.48479/f2a2-qw16} as Extended.csv, a sample of which is given in Table \ref{table:ExtendedCat}). These findings enrich the current catalog of GRGs in this region, offering insights into their distribution and contributing to an understanding of their growth mechanisms. This study underscores that even lower-bound redshift estimates can provide significant information on the physical properties of large radio galaxies.

\include{ExtendedSourceCat}
\begin{longrotatetable}
\begin{deluxetable*}{|c|c|c|c|c|c|c|c|}
\tablecaption{Extended radio sources with available NED redshifts}
\tablehead{\colhead{Position (J2000)} & \colhead{$\theta_{meas}$\tablenotemark{a}} & \colhead{$z_{NED}$\tablenotemark{b}} & \colhead{$z_{calc}$\tablenotemark{c}} & \colhead{$D_z$\tablenotemark{d}} &\colhead{$D_A$\tablenotemark{e}} & \colhead{$I_z$\tablenotemark{f}} &  \colhead{$I_A$\tablenotemark{g}} \\
\colhead{ } & \colhead{arcmin ($'$)} & \colhead{ } & \colhead{ } & \colhead{Mpc} &\colhead{Mpc} & \colhead{Mpc} &  \colhead{Mpc}
}
\startdata
\hline
17 12 42.79 -24 35 48.5 & 34.66 & 0.02433 S & 0.0229 & 106 & 90$\pm$40 & 1.067 & 0.911$\pm$0.399\\
17 09 28.15 -22 26 49.8 & 19.48 & 0.030955 S & 0.0388 & 133 & 150$\pm$62 & 0.752 & 0.85$\pm$0.35\\
16 57 30.02 -22 44 20.4 & 14.12 & 0.035164 P & 0.0415 & 150 & 160$\pm$70 & 0.617 & 0.658$\pm$0.288\\
17 17 02.70 -24 10 21.1 & 13.78 & 0.047319 S & 0.0492 & 201 & 188$\pm$78 & 0.804 & 0.754$\pm$0.311\\
17 05 00.60 -24 45 08.7 & 8.87 & 0.036292 S & 0.0538 & 153 & 205$\pm$84 & 0.396 & 0.528$\pm$0.218\\
17 11 37.56 -22 58 46.5 & 8.08 & 0.028666 S & 0.0412 & 123 & 159$\pm$65 & 0.288 & 0.373$\pm$0.154\\
19 00 31.63 -37 49 34.9 & 6.17 & 0.09352 P & 0.1296 & 368 & 451$\pm$186 & 0.662 & 0.81$\pm$0.334\\
16 49 51.74 -19 22 19.9 & 4.44 & 0.098692 S & 0.0914 & 403 & 332$\pm$137 & 0.52 & 0.43$\pm$0.177\\
16 51 32.39 -18 51 31.9 & 4.28 & 0.099109 S & 0.0749 & 410 & 278$\pm$114 & 0.511 & 0.346$\pm$0.143\\
18 29 07.32 -32 57 16.4 & 3.68 & 0.124246 S & 0.1036 & 501 & 371$\pm$153 & 0.537 & 0.398$\pm$0.164\\
18 58 36.96 -36 53 08.9 & 3.61 & 0.162561 P & 0.1255 & 643 & 439$\pm$181 & 0.675 & 0.461$\pm$0.19\\
16 37 26.55 -18 19 21.6 & 2.79 & 0.046928 P & 0.1154 & 187 & 408$\pm$179 & 0.152 & 0.332$\pm$0.145\\
16 36 19.20 -19 28 18.7 & 2.72 & 0.054137 S & 0.0584 & 228 & 221$\pm$97 & 0.18 & 0.175$\pm$0.076\\
16 43 01.88 -18 07 49.4 & 2.58 & 0.178622 P & 0.1153 & 713 & 408$\pm$168 & 0.535 & 0.306$\pm$0.126\\
16 45 12.68 -18 18 49.8 & 2.25 & 0.136392 P & 0.1093 & 547 & 389$\pm$171 & 0.359 & 0.255$\pm$0.112\\
17 14 09.72 -23 31 54.3 & 1.37 & 0.029307 S & 0.0441 & 125 & 170$\pm$70 & 0.05 & 0.068$\pm$0.028\\
16 48 45.09 -19 45 53.2 & 1.3 & 0.054901 S & 0.0689 & 229 & 257$\pm$106 & 0.087 & 0.097$\pm$0.04\\
18 48 47.93 -37 28 10.8 & 1.09 & 0.245472 P & 0.1173 & 978 & 414$\pm$181 & 0.309 & 0.131$\pm$0.057\\
17 11 45.12 -25 44 50.7 & 1.0 & 0.021131 S & 0.0559 & 89 & 212$\pm$87 & 0.026 & 0.062$\pm$0.025\\
18 34 42.42 -34 23 20.1 & 0.6 & 0.148643 P & 0.1324 & 584 & 459$\pm$189 & 0.102 & 0.08$\pm$0.033\\
18 55 55.84 -35 54 01.6 & 0.49 & 0.188393 P & 0.1059 & 758 & 378$\pm$166 & 0.108 & 0.054$\pm$0.024\\
16 47 39.46 -20 38 55.8 & 0.48 & 0.247823 P & 0.1609 & 950 & 541$\pm$223 & 0.131 & 0.075$\pm$0.031\\
\enddata
\tablenotetext{a}{The measured angular size of the radio source using OBIT software \citep{OBIT}.}
\tablenotetext{b}{The redshift obtained from the NASA Extragalactic Database (NED). S or P denotes spectroscopic or photometric.}
\tablenotetext{c}{The redshift estimated using K-band magnitudes.}
\tablenotetext{d}{The distance to the radio source calculated using $z_{NED}$.}
\tablenotetext{e}{The distance to the radio source calculated using $z_{calc}$.}
\tablenotetext{f}{The projected physical size of the radio source derived using $D_z$.}
\tablenotetext{g}{The projected physical size of the radio source derived using $D_A$.}
\label{table:ExtendedCat}
\end{deluxetable*}
\end{longrotatetable}

\begin{longrotatetable}
\begin{deluxetable*}{|c|c|c|c|c|c|c|c|}
\tablecaption{Probable Giant Radio Galaxies}
\tablehead{\colhead{Position (J2000)} & \colhead{$\theta_{meas}$\tablenotemark{a}} & \colhead{$z_{NED}$\tablenotemark{b}} & \colhead{$z_{calc}$\tablenotemark{c}} & \colhead{$D_z$\tablenotemark{d}} &\colhead{$D_A$\tablenotemark{e}} & \colhead{$I_z$\tablenotemark{f}} &  \colhead{$I_A$\tablenotemark{g}} \\
\colhead{ } & \colhead{arcmin (')} & \colhead{ } & \colhead{ } & \colhead{$Mpc$} &\colhead{$Mpc$} & \colhead{$Mpc$} &  \colhead{$Mpc$}
}
\startdata
\hline
17 12 42.79 -24 35 48.5 & 34.66 & 0.02433 S & 0.0229 & 106 & 90$\pm$40 & 1.067 & 0.911$\pm$0.399\\
17 09 28.15 -22 26 49.8 & 19.48 & 0.030955 S & 0.0388 & 133 & 150$\pm$62 & 0.752 & 0.85$\pm$0.35\\
17 17 02.70 -24 10 21.1 & 13.78 & 0.047319 S & 0.0492 & 201 & 188$\pm$78 & 0.804 & 0.754$\pm$0.311\\
18 04 10.37 -34 21 07.2& 12.88 &   & 0.0914 &   & 333$\pm$137 &   & 1.246$\pm$0.513\\
18 12 43.18 -31 50 04.5 & 12.73 &   & 0.0705 &   & 263$\pm$108 &   & 0.973$\pm$0.401\\
17 28 08.76 -22 39 40.5 & 9.31 &   & 0.09 &   & 328$\pm$135 &   & 0.887$\pm$0.366\\
18 29 01.82 -34 14 37.18 & 6.44 &   & 0.1107 &   & 394$\pm$162 &   & 0.738$\pm$0.304\\
17 05 47.14 -22 56 47.4 & 6.28 &   & 0.122 &   & 428$\pm$177 &   & 0.783$\pm$0.323\\
19 00 31.63 -37 49 34.9 & 6.17 & 0.09352 P & 0.1296 & 368 & 451$\pm$186 & 0.662 & 0.81$\pm$0.334\\
16 50 41.76 -22 04 18.5 & 5.22 &   & 0.2571 &   & 779$\pm$341 &   & 1.183$\pm$0.518\\
16 43 27.97 -17 21 37.7 & 5.12 &   & 0.2397 &   & 740$\pm$324 &   & 1.102$\pm$0.483\\
16 57 09.35 -20 22 28.6 & 4.26 &   & 0.2194 &   & 692$\pm$303 &   & 0.858$\pm$0.376\\
16 41 20.55 -18 11 38.9 & 3.95 &   & 0.5501 &   & 1253$\pm$549 &   & 1.438$\pm$0.63\\
18 25 04.80 -33 53 59.0 & 3.71 &   & 0.2299 &   & 717$\pm$296 &   & 0.773$\pm$0.319\\
18 20 050.85 -32 53 25.8 & 3.56 &   & 0.3592 &   & 981$\pm$405 &   & 1.017$\pm$0.419\\
16 39 54.80 -19 08 32.0 & 3.17 &   & 0.2826 &   & 834$\pm$365 &   & 0.77$\pm$0.337\\
16 45 48.67 -19 25 43.9 & 2.73 &   & 0.5008 &   & 1193$\pm$523 &   & 0.948$\pm$0.415\\
\enddata
\tablenotetext{a}{The measured angular size of the radio source using OBIT software \citep{OBIT}.}
\tablenotetext{b}{The redshift obtained from the NASA Extragalactic Database (NED).  S or P denotes spectroscopic or photometric.}
\tablenotetext{c}{The redshift estimated using K-band magnitudes.}
\tablenotetext{d}{The distance to the radio source calculated using $z_{NED}$.}
\tablenotetext{e}{The distance to the radio source calculated using $z_{calc}$.}
\tablenotetext{f}{The projected physical size of the radio source derived using $D_z$.}
\tablenotetext{g}{The projected physical size of the radio source derived using $D_A$.}
\label{table:GRG_list}
\end{deluxetable*}
\end{longrotatetable}

\section{Summary\label{Summary}}
We present a MeerKAT survey of the Bulge region of the Milky Way with 8\asec\ resolution at 1.3 GHz in full polarization with a combination of completely covered strips above and below the Galactic Center as well as samples of fields at low longitudes near the Galactic plane.
Most of the Stokes I emission detected is from the extragalactic background sky.
Displays of especially well resolved background AGNs are presented.
Widespread linearly polarized emission is seen from the Faraday rotation of the polarized Galactic disk emission by fine scale magnetized structure in the ISM.
Catalogs of relatively compact sources are given to help identify Galactic objects;  especially polarized sources are strongly biased towards Galactic stars.
A search for Giant Radio Galaxies (GRG) identifies 17 probabl\chg{e} new GRGs.

\section*{Acknowledgments}
\chg{We thank the anonymous reviewer for insightful comments leading to an improved presentation.}
The MeerKAT telescope is operated by the South African Radio Astronomy
Observatory, which is a facility of the National Research Foundation,
an agency of the Department of Science and Innovation. 
The National Radio Astronomy Observatory is a facility of the National
Science Foundation, operated under a cooperative agreement by
Associated Universities, Inc. 
This research has made use of the SIMBAD database,
operated at CDS, Strasbourg, France.
This research has made use of the NASA/IPAC Extragalactic Database,
which is funded by the National Aeronautics and Space Administration
and operated by the California Institute of Technology.

\vspace{5mm}
\facilities{MeerKAT}

\software{Obit \cite{OBIT}}
\bibliography{Bulge}{}
\bibliographystyle{aasjournal}


\end{document}